\documentclass[conference]{IEEEtran}
\IEEEoverridecommandlockouts

\usepackage{cite}
\usepackage{amsmath,amssymb,amsfonts,bm}
\usepackage{graphicx}
\usepackage{textcomp}
\usepackage{xcolor}
\usepackage{comment}
\usepackage{amsmath,amssymb,amsfonts}
\usepackage{pgfplots}
\pgfplotsset{compat=newest}
\usepgfplotslibrary{groupplots} 
\usepackage{tikz}
\usetikzlibrary{arrows,automata,positioning}
\usetikzlibrary{arrows.meta}
\usepackage{graphicx}
\usepackage{textcomp}
\usepackage{xcolor}
\usepackage{lipsum}
\usepackage{booktabs}
\usepackage{color, colortbl}
\usepackage{algorithm}
\usepackage{algpseudocode}
\usepackage{dsfont}
\usepackage[acronyms]{glossaries}
\usepackage{tcolorbox}
\usetikzlibrary{calc, positioning, arrows.meta, fit}
\usetikzlibrary{svg.path}

\def\BibTeX{{\rm B\kern-.05em{\sc i\kern-.025em b}\kern-.08em
    T\kern-.1667em\lower.7ex\hbox{E}\kern-.125emX}}

\pgfdeclarelayer{lay00}
\pgfdeclarelayer{back00}
\pgfdeclarelayer{lay01}
\pgfdeclarelayer{back01}
\pgfdeclarelayer{lay02}
\pgfdeclarelayer{back02}
\pgfdeclarelayer{lay03}
\pgfdeclarelayer{back03}
\pgfdeclarelayer{back04}
\pgfdeclarelayer{lay10}
\pgfdeclarelayer{back10}
\pgfdeclarelayer{lay11}
\pgfdeclarelayer{back11}
\pgfdeclarelayer{lay12}
\pgfdeclarelayer{back12}
\pgfdeclarelayer{lay13}
\pgfdeclarelayer{back13}
\pgfdeclarelayer{back14}
\pgfdeclarelayer{lay30}
\pgfdeclarelayer{back30}
\pgfdeclarelayer{lay31}
\pgfdeclarelayer{back31}
\pgfdeclarelayer{lay32}
\pgfdeclarelayer{back32}
\pgfdeclarelayer{lay33}
\pgfdeclarelayer{back33}
\pgfdeclarelayer{back34}

\pgfsetlayers{back34,back33,lay33,back32,lay32,back31,lay31,back30,lay30,back14,back13,lay13,back12,lay12,back11,lay11,back10,lay10,back04,back03,lay03,back02,lay02,back01,lay01,back00,lay00,main}

\newlength{\figSpaceX}
\newlength{\figSpaceY}
\newlength{\nodeSize}
\colorlet{group color}{white}
\colorlet{components color}{black!15!white}

\newacronym{rcs}{RCS}{radar cross section}
\newacronym{cfar}{CFAR}{constant false alarm rate}
\newacronym{dtt}{DTT}{detect-then-track}
\newacronym{mot}{MOT}{multiobject tracking}
\newacronym{tbd}{TBD}{track-before-detect}
\newacronym{rfs}{RFS}{random finite set}
\newacronym{bp}{BP}{belief propagation}
\newacronym{pdf}{pdf}{probability density function}
\newacronym{snr}{SNR}{signal-to-noise ratio}
\newacronym{sbl}{SBL}{sparse Bayesian learning}
\newacronym{ospa}{OSPA}{optimal sub-pattern assignment}
\newacronym{vmp}{VMP}{variational message passing}
\newacronym{slam}{SLAM}{simultaneous localization and mapping}
\newacronym{po}{PO}{potential objects}
\newacronym{elbo}{ELBO}{evidence lower bound}

\newcommand{\trans}{\text{T}}

\newcommand{\diag}{\text{diag}}

\newcommand{\ist}{\hspace*{.3mm}}
\newcommand{\rmv}{\hspace*{-.3mm}}
\newcommand{\iist}{\hspace*{1mm}}

\newcommand{\nn}{\nonumber}

\colorlet{plot1}{cyan!40!white}
\colorlet{plot2}{red!20!white}
\colorlet{plot3}{cyan!70!black}
\colorlet{plot4}{red!40!gray}
\colorlet{plot5}{red}

\newlength{\figurewidth}
\newlength{\figureheight}

\begin{document}

\title{\huge
A Variational Message Passing Framework for Multi-Sensor Multi-Object Tracking using Raw Radar Signals
\\

\thanks{
This work has received funding in part by the Thomas B. Thriges Foundation grant 7538-1806 and in part by the Austrian Research Promotion Agency (FFG) under the PRISM project (620753).}
}

\author{
    \IEEEauthorblockN{Anders Malthe Westerkam$^\dagger$, Jakob Möderl$^*$, Erik Leitinger$^*$, and Troels Pedersen$^\dagger$
    }
    \IEEEauthorblockA{$\dagger$Aalborg University, Aalborg Denmark. Email: \{amw troels\}@es.aau.dk}
    \IEEEauthorblockA{$*$Graz University of Technology, Graz, Austria. Email: \{jakob.moderl, erik.leitinger\}@tugraz.at}
    \vspace{-5mm}
}

\maketitle

\begin{abstract}

The growing proliferation of unmanned aerial vehicles (UAVs) poses major challenges for reliable airspace surveillance, as drones are typically small, have low radar cross-sections, and often move slowly in cluttered environments. These characteristics make the joint tasks of detecting, localizing, and tracking multiple objects difficult for conventional \gls{dtt} approaches, which rely on pre-processed measurements and may discard informative low-\gls{snr} signal components. To overcome these limitations, we propose a \gls{vmp}-based direct \gls{mot} method that operates directly on raw radar signals and explicitly accounts for an unknown and time-varying number of objects.

The proposed method is formulated for MIMO multi-radar systems and performs data fusion by jointly processing the signals of all radar sensors using a probabilistic model. A superimposed signal model is employed to capture correlations in the raw sensor data caused by closely spaced objects, and a hierarchical Bernoulli-Gamma model is introduced to jointly model object existence, reflectivities, and the reliability of individual radar-object links. Using a mean-field approximation, we derive message updates, yielding a computationally efficient \gls{vmp} algorithm that simultaneously performs object detection, track formation, state estimation, and nuisance parameter learning directly from the radar signal. Simulation results in synthetic scenarios with weak and closely-spaced objects show that the proposed direct-\gls{mot} method outperforms a conventional pipeline based on super-resolution estimation followed by \gls{bp}-based tracking, particularly in low-\gls{snr} and clutter-rich conditions, demonstrating the advantages of direct signal-level inference and coherent multi-radar fusion.

\end{abstract}

\glsresetall
\section{Introduction}

In recent years, unmanned aerial vehicles (UAVs), or drones, have become increasingly prevalent in private, commercial, and military applications. Their decreasing cost and the ability to operate multiple drones simultaneously have lowered the barrier to accessing controlled airspace, increasing the risk of accidental or intentional intrusions into restricted areas such as airports or critical infrastructure. Radar systems are well-suited for monitoring such airspace due to their robustness to lighting and weather conditions \cite{Richards2014}. However, radar-based drone detection is challenging because drones are typically small, exhibit low \gls{rcs}, and often move slowly relative to surrounding clutter. As a result, conventional Doppler-based high-pass filtering may fail to reliably separate drone signatures from background clutter \cite{Poitevin2017,Gong2023,Quevedo2019}.

These characteristics lead to a sensing problem that inherently involves detecting, localizing, and tracking multiple weak and slow-moving objects in cluttered environments. More generally, this task falls within the scope of \gls{mot}, where the time-varying states of multiple objects must be inferred from sensor data under uncertainty regarding the number of objects, their appearance and disappearance, and their close spatial proximity. Such conditions introduce challenges including missed detections, clutter, and reliable track management (initialization and termination). Therefore, robust \gls{mot} methods that can operate under low-\gls{snr} and clutter-rich conditions are essential for drone surveillance and airspace protection.

\subsection{State-of-the-Art}

In radar signal processing, \gls{mot}, i.e., the joint tasks of detecting, localizing, and tracking multiple objects has traditionally been addressed using \gls{dtt} algorithms that operate on pre-processed object estimates (measurements) rather than raw radar signals \cite{Wei2017}. In these approaches, a front-end detector first extracts measurements (e.g., matched filtering and \gls{cfar} detection \cite{RobFuhKelNit:TAES1992}), which are then used by a tracker to infer object states from one or multiple sensors, even when the number of objects is unknown \cite{BarWilTia:B11, Mahler2007, MeyerProc2018, LiLeiVenTuf:TWC2022, VenLeiTerMeyWit:TWC2024, Zhang2024}. While \gls{dtt}-based \gls{mot} methods are computationally efficient and widely adopted, their performance is fundamentally limited by the quality of the intermediate detections. In particular, weak objects can be masked by clutter or noise \cite{Ristic2020}, leading to missed detections and the loss of informative signal energy before the tracking stage. 

To mitigate this limitation, \gls{tbd} methods have been developed that operate directly on raw radar signals instead of intermediate measurements, thereby preserving low-\gls{snr} information and improving performance in weak-object scenarios \cite{TonSha:TAES1998, Rutten2005, Davey2007, MoySpaLam:TAES2011, PapVoVoFanBea:TAES2015, LepRabLeG:TAES2016, Ristic2020, Lehmann2012, Zhichao2020}. Early \gls{tbd} approaches include batch-processing techniques based on maximum likelihood estimation \cite{TonSha:TAES1998}, the Hough transform \cite{MoySpaLam:TAES2011}, and dynamic programming \cite{Barniv:TAES1985}; however, their computational burden is typically prohibitive for real-time applications. More recent real-time Bayesian \gls{tbd} \gls{mot} methods for an unknown number of objects can broadly be categorized into \gls{rfs}-based filters \cite{Kropfreiter2024, Ristic2020, KimRisGuaRos:TAES2021} and factor-graph-based methods using \gls{bp} \cite{Liang2023}, which exploit the natural factorization of the posterior \gls{pdf} to achieve scalability. For instance, the \gls{bp}-based approach in \cite{Liang2023} employs a spatial cell grid whose cells may contain objects or noise, accounts for the fact that a single object can contribute to multiple cells, and incorporates a birth-death model to enable automatic track initiation and termination, demonstrating robustness in low-\gls{snr} regimes.

Despite these advances, most existing \gls{tbd} methods rely on simplified likelihood models to approximate the radar signal. Common approximations include the use of point-spread function models, known nuisance parameters (e.g., amplitudes and noise variance) \cite{Ristic2020,PapVoVoFanBea:TAES2015}, matched-filtered signals as sufficient statistics \cite{Lehmann2012}, and separable likelihoods in which samples are treated as independent and at most one object is assumed per sample \cite{Ristic2020,KroWilMey:FUSION2021,Kropfreiter2024}. These approximations neglect signal correlations induced by superposition and closely spaced objects and can therefore lead to performance degradation in dense or low-\gls{snr} scenarios. More general likelihood formulations that directly incorporate the radar signal have been investigated in \cite{LepRabLeG:TAES2016}. Building on this line of work, the direct multipath-based \gls{slam} framework in \cite{LiaLeiMey:Asilomar2023,LiaLeiMey:TSP2025}, which is closely related to \gls{mot}, employs \gls{bp} message passing for superposition measurement models and operates directly on the full radar signal. The same principle are applied to \gls{mot} in \cite{LiaMey:Asilomar2024, LiaMey:Arxiv2025}. As a direct approach \cite{BiaRapWei:TVT2013}, it jointly handles unknown nuisance parameters (e.g., amplitudes, object \gls{snr}, and noise variance) and explicitly models correlations in the raw sensor signal caused by multiple closely spaced objects.

Alternative message-passing-based \gls{mot} methods adopt a variational Bayesian formulation, i.e., \gls{vmp}, but typically still follow a \gls{dtt} paradigm and therefore rely on pre-processed measurements \cite{LundgrenTSP2016,GanLiGod:TAES2024,BaiLanWanPanHaoLi:TAES2024}. A prominent real-time variational \gls{tbd} \gls{mot} method is the Histogram Probabilistic Multi-Hypothesis Tracker (HPMHT) \cite{Davey2007,DavWieVu:JSTSP2013}, which is based on an expectation-maximization formulation; however, its performance is sensitive to parameter tuning and model specification \cite{KimRisGuaRos:TAES2021}. 
In \cite{Westerkam2023}, a \gls{vmp}-based algorithm operating directly on the radar signal was proposed for tracking a single low-\gls{snr} object.

\subsection{Contributions and Paper Organization}

In this paper, we present a \gls{vmp}-based direct-\gls{mot} method that accounts for an unknown number of objects and integrates both object detection and track formation for MIMO multi-radar systems. Specifically, our method employs a mean-field approximation and, in accordance with \cite{LiaLeiMey:Asilomar2023, LiaLeiMey:TSP2025}, is built on a superimposed signal model that captures correlations in raw sensor signals caused by closely spaced objects. In contrast to single-sensor formulations, we explicitly perform data fusion across multiple MIMO radars by jointly processing the matched-filtered signals of all radar nodes within a unified probabilistic model. Inspired by \cite{Badiu2017, Moederl2024}, this formulation enables the simultaneous estimation of object existence, modeled by a binary random variable, and individual object states (position, velocity, and potentially other kinematic parameters). Moreover, we adopt a hierarchical Bernoulli-Gamma model in which the Bernoulli variables describe the existence of individual objects, while Gamma-distributed precision parameters govern the reflectivities, i.e., amplitude precision (inverse variance), and implicitly capture the reliability of each object-radar link in the multi-radar fusion process. This allows the method to inherently consider the impact of unreliable sensing links and fuse heterogeneous MIMO radar observations robustly within a Bayesian framework. 

The main contributions of this paper are as follows.
\begin{itemize}
    \item  We introduce a novel direct-\gls{mot} method based on \gls{vmp} that jointly estimates the number of objects and their individual states while performing coherent data fusion across multiple MIMO radars.
    \item We incorporate a Bernoulli-Gamma hierarchical model that simultaneously captures object existence and the reliability of individual radar-object links, enabling adaptive weighting of multi-radar measurements.
    \item We derive \gls{vmp} message updates that consider correlations within the radar signals and remain computationally efficient by exploiting a mean-field approximation.
    \item We demonstrate that our method outperforms a \gls{dtt} approach consisting of a super-resolution \gls{sbl}-based estimation stage \cite{HansenSAM2014,GreLeiWitFle:TWC2024,Moederl2024,MoeWesVenLei:Fusion2025}  followed by \gls{bp}-based \gls{mot} \cite{MeyerProc2018,LiLeiVenTuf:TWC2022}.
\end{itemize}
This paper advances the authors' current line of research \cite{WesMoeLeiPed:Fusion2025} by (i) we extend the statistical model by generalizing the amplitude model and introducing precision parameters that are inferred individually for each object--radar link, (ii) we extend the framework to MIMO multi-radar systems with joint data fusion based on the proposed statistical model (hierarchical Bernoulli-Gamma model), (iii) we provide a comprehensive description of the underlying statistical model, (iv) we derive all mean-field message updates and the resulting \gls{vmp}-based method in detail, (v) we provide an analysis of the algorithms' computational complexity, and (vi) we present extensive experimental evaluations in synthetic radar tracking scenarios.

The remainder of this paper is organized as follows. In Sec.~\ref{sec:system_model}, introduces the MIMO radar signal model and define the underlying physical measurement model based on superimposed raw sensor signals. Sec.~\ref{sec:radsignal} presents the probabilistic system model, including the state evolution, existence indicators, reflectivity, and hierarchical prior modeling, and derives the joint posterior \gls{pdf}. In Sec.~\ref{sec:alg_construction}, we develop the proposed \gls{vmp}-based direct multiobject tracking algorithm, including the mean-field approximation, surrogate \glspl{pdf}, and practical implementation aspects for multi-radar data fusion. Sec.~\ref{sec:numSim} provides numerical simulations and performance evaluations in challenging low-\gls{snr} and closely-spaced object scenarios. Finally, Sec.~VI concludes the paper.

\section{MIMO Radar Signal Model}\label{sec:radsignal}
\begin{figure*}
    \setlength{\figurewidth}{0.8\linewidth}
    \setlength{\figureheight}{3cm}
    \centering
    \scalebox{0.65}{
    \input{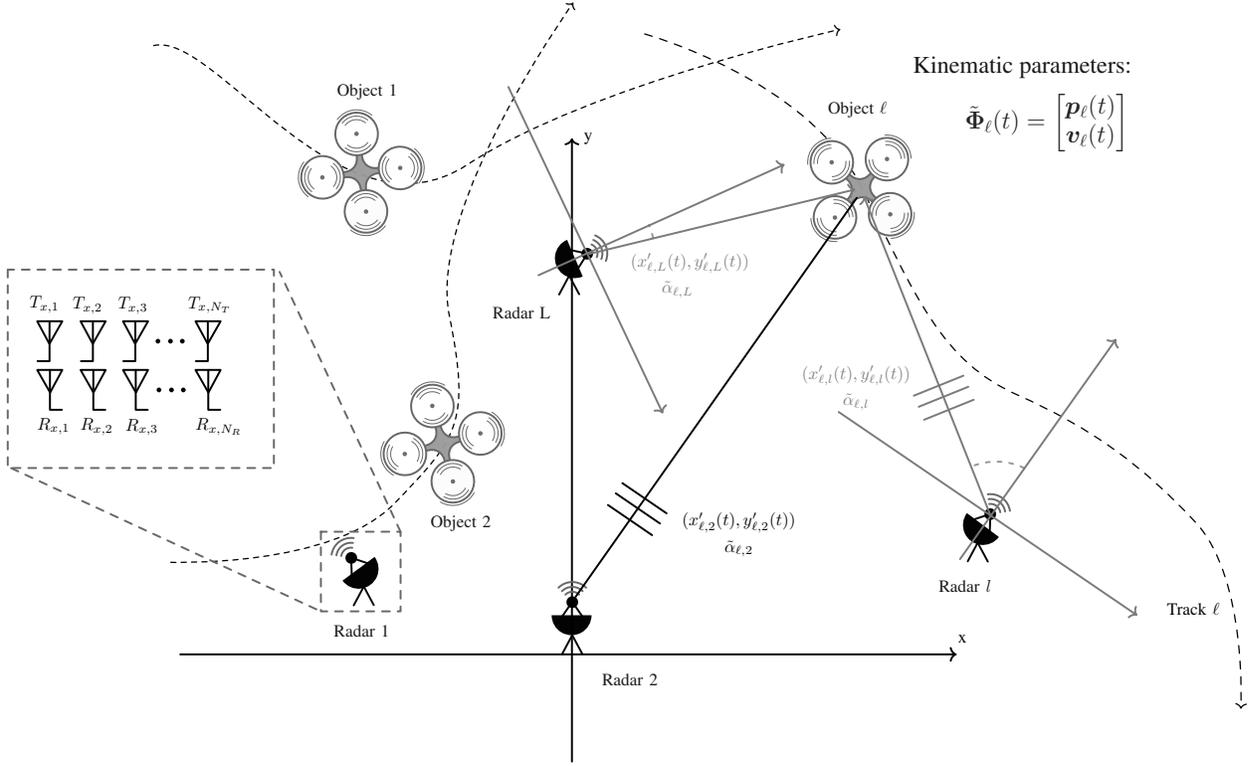}
    }
    \vspace*{-17mm}
    \caption{Scenario with an unknown number of objects $\mathcal{L}(t)$, each with their own kinematic parameters $\tilde{\bm{\Phi}}_{\ell}(t)$ being observed by $L$ $N_T\times N_R$ MIMO radars, resulting in complex amplitudes $\tilde{\alpha}_{\ell,l}(\tau)$ per object and radar.}
    \label{fig:Overveiw}
\end{figure*}

Consider a scenario as depicted in Fig.~\ref{fig:Overveiw} with $\mathcal{L}(t)$ objects at time $t$ in a clutter-free environment. Referring to a coordinate system, each object $\ell=1,\dots,\mathcal{L}(t)$ is characterized by its state-space parameters $\tilde{\bm{\Phi}}_{\ell}(t) = [\bm{p}_{\ell}(t)^\trans \iist \bm{v}_{\ell}(t)^\trans]^\trans$, describing its 2D position $\bm{p}_{\ell}(t) = [\tilde{x}_{\ell}(t)\iist \tilde{y}_{\ell}(t)]^\trans$ and velocity $\bm{v}_{\ell}(t) = [\tilde{v}_{\text{x},\ell}(t)\iist \tilde{v}_{\text{y},\ell}(t)]^\trans$, and by its radar cross section (RCS) $\sigma_{\text{R},l,\ell}$. The problem considered is to estimate both the number of objects  $\mathcal{L}(t)$, and their respective states $\tilde{\bm{\Phi}}_{\ell}(t)$ based on all received data before time $t$.

The radar system consists of $L$ monostatic MIMO radars each equipped with $N_T$ transmitters and $N_R$ receivers each antenna is assumed isotropic. Each transmitter $m$ emits a baseband signal $u_m(t)$ at carrier frequency $f_{\mathrm{c},l}$, signals from different transmitters are assumed mutually orthogonal, and the radars are assumed non interfering. We denote a MIMO pulse as the transmission of all transmitters, with transmission time $T_{\mathrm{MIMO}}$. The time between consecutive MIMO pulses is $\Delta t$, giving a pulse repetition frequency (PRF) $\mathrm{PRF} = 1/\Delta t$. All objects in the radar’s field of view (FOV) reflect each pulse. Assuming slow object motion relative to the PRF, a “stop-and-hop” model is used:  the number of objects and the kinematic parameters of the $\ell$-th object remain constant during the $n$-th interval, i.e., $\tilde{\bm{\Phi}}_{\ell}(t) = \tilde{\bm{\Phi}}_{\ell,n}$ and $\mathcal{L}(t)=\mathcal{L}_n$ for $n \Delta t \leq t \leq (n+1)\Delta t$. Since velocities are low, Doppler effects are neglected. 
Only the direct path is considered, and all objects lie in the far field. After down-conversion, the baseband signal at the $j$-th receiver on the $l$ radar is 
\vspace*{-2mm}
\begin{align}
    y_{j,n,l}(\tau) &= \sum_{\ell=1}^{\mathcal{L}_n}\sum_{m=1}^{N_T}\tilde{\alpha}_{\ell,n,l}a_{j,m,\ell,l}(\tilde{\theta}_{\ell,n,l})\nn\\[-2mm] 
    &\hspace{18mm} \times u(\tau-\tilde{\tau}_{\ell,n,l})+w_{j,l}(\tau)\label{eq:signal_model_time}\\[-7mm]\nn
\end{align}
where $\tilde{\alpha}_{\ell,n,l} = |\tilde{\alpha}_{\ell,n,l}|e^{i\,(2\pi\,f_{\mathrm{c}}\,\tilde{\tau}_{\ell,n,l} + \phi_{\ell,n,l})}$ is the complex weight of the $\ell$-th object seen from the $l$-th radar with magnitude $|\tilde{\alpha}_{\ell,n,l}| = \sqrt{\sigma_{\mathrm{R},\ell,l}} \frac{f_{\text{c},l}}{(4\pi)^{3/2}( \tau_{(\ell,n,l)})^2}$ and random phase $\phi_{\ell,n,l}$. The magnitude comprises the object-dependent radar cross section $\sigma_{\mathrm{R},\ell,l}$ and a path-loss-related term. The array steering function $a_{j,m,l}(\tilde{\theta}_{\ell,n,l})$ at receiver $j$ for transmitter $m$ is evaluated at the bearing $\tilde{\theta}_{\ell,n,l} = \arctan(x'_{\ell,n,l} / y'_{\ell,n,l})$ seen from the local coordinate system of radar $l$ shown in Fig.~\ref{fig:Overveiw}. The term $u_m\bigl(\tau - \tilde{\tau}_{\ell,n,l}\bigr)$ is the transmitted baseband signal from transmitter $m$ delayed by $\tilde{\tau}_{\ell,n,l}$, where $\tilde{\tau}_{\ell,n,l} = 2\,\sqrt{(x'_{\ell,n,l})^2 + (y'_{\ell,n,l})^2} / c$ represents the two-way propagation time for the $\ell$-th object with $c$ as the speed of light. The last term $w_j(\tau)$ denotes white, complex, circularly-symmetric Gaussian noise with variance $\sigma_w^2$.

\section{System Model}\label{sec:system_model}

\subsection{State Vectors and Inference Signal Model}

We consider $K$ \gls{po}  with $K > \mathcal{L}_n$ and time-varying states.
The $k$-th PO at time $n$ has state $\bm{\Phi}_{k,n} = [\bm{p}^\trans_{k,n} \iist \bm{v}^\trans_{k,n}]^\trans$ and complex-valued reflectivity $\alpha_{k,n,l}$, which also incorporates the path-loss term. Each PO is also associated with a binary existence indicator $\xi_{k,n}$ taking value one if the \gls{po} is present and zero otherwise. Then, the number of \glspl{po} at time $n$ then reads
$\mathcal{L}_n=\sum_{k=1}^{K}\xi_{n,k}$. After sampling and matched filtering in the frequency domain, the received signal at radar $l$, denoted by $\bm{Z}_{n,l}$, can be expressed as
\vspace*{-2mm}
\begin{align}\label{eq:Z_constructio}
    \bm{Z}_{n,l} 
    = \sum_{k=1}^{K} \alpha_{k,n,l}\, \xi_{n,k} \, \bm{S}_l\bigl(\bm{\Phi}_{n,k}\bigr) + \tilde{\bm{W}}_l \qquad \in \mathbb{C}^{N_Z \times 1}
    \\[-7.5mm]\nonumber
\end{align}
where $N_Z = N_\text{s} N_\text{T} N_\text{R}$, and $N_\text{s}$ denotes the number of time (or frequency) samples obtained from a single radar pulse. Note that the maximum number of \glspl{po} is given by the signal samples, i.e., $K_\text{max}=N_Z$. Here $\tilde{\bm{W}}_{l} \in \mathbb{C}^{N_Z \times 1}$ is a zero-mean circularly-symmetric complex AWGN with covariance matrix $\bm{\Lambda}_Z$. The spatiotemporal steering vector from the $k$-th PO at time $n$ radar $l$ is given by
\vspace{-2mm}
\begin{align}
   \bm{S}_l\bigl(\bm{\Phi}_{n,k}\bigr) = \sum_{m=1}^{N_\text{T}} \bm{A}_{m,l}\bigl(\theta_{n,k,l}\bigr)\otimes \bm{h}(\tau_{n,k,l})\\[-7.5mm] \nn
\end{align}
Here, $\bm{A}_{m,l}(\theta_{n,k,l}) \in \mathbb{C}^{N_\text{R} \times 1}$ is the receive steering vector for bearing $\theta_{n,k}$, $\bm{h}(\tau_{n,k,l})\in\mathbb{C}^{N_\text{T} N_s \times 1}$ is the matched-filtered transmit-signal spectrum for delay $\tau_{n,k,l}$, and $\otimes$ denotes the Kronecker product. 

\subsection{Probabilistic Model}
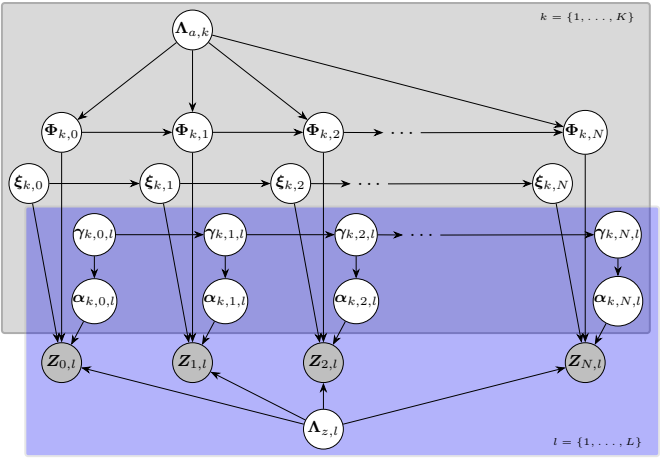
\begin{figure}[t]
    \centering
    \resizebox{\columnwidth}{!}{%
    \begin{tikzpicture}[variable node/.style={circle,yshift=0.2,draw,minimum width=\nodeSize,inner sep=0,font=\footnotesize,fill=white},
	deterministic node/.style={circle,dashed,draw,minimum width=\nodeSize,font=\footnotesize,fill=white}, data node/.style={circle,yshift=0.2,draw,minimum width=\nodeSize,inner sep=0,font=\footnotesize,fill=lightgray},
	pdf/.style={draw,fill=white,font=\scriptsize,rounded corners=2pt},
	scale=1,>={Stealth[scale=1]}
	]
\node[variable node] at (1\figSpaceX,3\figSpaceY) (Lambdaa){$\boldsymbol{\Lambda}_{a,k}$};

\node[variable node] at (2\figSpaceX,-4.8\figSpaceY) (L){$\boldsymbol{\Lambda}_{z,l}$};

\foreach \j in {0,1,2,3,4}{

    \ifnum \j = 3
    \pgfmathparse{(\j-0.4)}
    \node at (\pgfmathresult\figSpaceX,1\figSpaceY) (PHI\j){$\hdots$};
    \pgfmathparse{(\j-0.65)};
    \node at (\pgfmathresult\figSpaceX,0\figSpaceY) (xi\j){$\hdots$};    
    \pgfmathparse{(\j-0.25)};
    \node at (\pgfmathresult\figSpaceX,-1\figSpaceY) (gamma\j){$\hdots$};    
    \else
    \ifnum \j = 4
    \def \lval{N};
    \else
    \def \lval{\j};
    \fi
    \pgfmathparse{(\j-0)}
    \node[variable node] at (\pgfmathresult\figSpaceX,1\figSpaceY) (PHI\j){$\boldsymbol{\Phi}_{k,\lval}$};
    \pgfmathparse{(\j+0)}; 
    \node[data node] at (\pgfmathresult\figSpaceX,-3.5\figSpaceY) (Z\j){$\boldsymbol{Z}_{\lval,l}$};
    \pgfmathparse{(\j-0.25)};
    
    \node[variable node] at (\pgfmathresult\figSpaceX,0\figSpaceY) (xi\j){$\boldsymbol{\xi}_{k,\lval}$};
    \pgfmathparse{(\j+0.25)}; 
    
    \node[variable node] at (\pgfmathresult\figSpaceX,-2.3\figSpaceY) (alpha\j){$\boldsymbol{\alpha}_{k,\lval,l}$};

    \node[variable node] at (\pgfmathresult\figSpaceX,-1.0\figSpaceY) (gamma\j){$\boldsymbol{\gamma}_{k,\lval,l}$};
    
    \draw[->] (Lambdaa)--(PHI\j);
    \draw[->] (PHI\j)--(Z\j);
    \draw[->] (xi\j)--(Z\j);
    \draw[->] (alpha\j)--(Z\j);
    \draw[->] (L)--(Z\j);
    \draw[->] (gamma\j)--(alpha\j);

    \fi
    
    \ifnum \j > 0
    \pgfmathparse{int(\j-1)};
    \draw[->] (PHI\pgfmathresult)--(PHI\j);
    \draw[->] (xi\pgfmathresult)--(xi\j);
    \draw[->] (gamma\pgfmathresult)--(gamma\j);
    \fi
};
\node at (-0.80, -1) (dummy){};
\begin{pgfonlayer}{back01}
                \node[fill=components color,draw=components color!80!black, rounded corners=0.5mm, line width=1pt,fit=(xi0) (PHI4) (Lambdaa) (alpha4)] (0pt,0pt) {};
                \node[anchor=base west] at (3.6\figSpaceX,3.2\figSpaceY) {\tiny $k=\{1,\hdots,K\}$};
\end{pgfonlayer}
\begin{pgfonlayer}{back01}
\node[
    fill=black!5!blue,
    opacity=0.3,
    draw=components color!80!black,
    rounded corners=0.5mm,
    line width=1pt,
    inner xsep=8pt,          
    fit=(Z0) (gamma0) (alpha4) (L)
] {};
\node[anchor=base west] at (3.7\figSpaceX,-5.1\figSpaceY) {\tiny $l=\{1,\hdots,L\}$};
\end{pgfonlayer}


\end{tikzpicture}
}
    \vspace{-4mm}
    \caption{%
      Bayesian network representation of the multi-object detection and tracking problem.
      Each object $k$ has state $\bm{\Phi}_{k,n}$, existence $\xi_{k,n}$, and reflectivity $\alpha_{k,n,l}$. 
      Shaded nodes denote the measured data $\bm{Z}_n$ and clear nodes represent unknown random variables.
      }
    \label{fig:Baysian_graph}
\end{figure}

The signal model is now complemented with a probabilistic model resulting in the Bayesian network shown in Fig.~\ref{fig:Baysian_graph}.
The existence indicator $\xi_{k,n}$, the object state $\bm{\Phi}_{k,n}$, and complex-valued reflectivities $\alpha_{k,n,l}$, process-noise covariances matrices $\bm{\Lambda}_{k,\text{a}}$ are considered unknown and time-varying. To ease notation we define the stacked vectors 
$\bm{\Phi}_n \triangleq [\bm{\Phi}_{1,n}^{\trans}, \dots, \bm{\Phi}_{K,n}^{\trans}]^{\trans}$,
$\bm{\xi}_n \triangleq [\xi_{1,n}, \dots, \xi_{K,n}]^{\trans}$,
$\bm{\alpha}_{n,l} \triangleq [\,\alpha_{1,n,l},\ist \cdots, \ist\,\alpha_{K,n,l}\,]^{\trans}$. Similarly, let 
$\bm{\Phi}_{0:N} \triangleq [\bm{\Phi}_0,\ist \cdots, \ist\bm{\Phi}_N]$,
$\bm{\xi}_{0:N} \triangleq [\bm{\xi}_0,\ist \cdots, \ist\bm{\xi}_N]$,
$\bm{\alpha}_{0:N,l} \triangleq [\bm{\alpha}_{0,l},\ist \cdots, \ist\bm{\alpha}_{N,l}]$, and
$\bm{Z}_{0:N,l} \triangleq [\bm{Z}_{0,l},\ist \cdots, \ist\bm{Z}_{N,l}]$ denote the stacked vectors over all timestamps $n=0,\dots,N$, where $N$ refers to the last recorded time instance, and hence grow as more data is collected.

\subsubsection{State Transition Model}

We assume that the \gls{po} state evolve independently across $k$ and $n$ and the joint state-transition \gls{pdf} factorizes as 
\vspace*{-2mm}
\begin{align}\label{eq:SST}
    p(\bm{\Phi}_{n} \mid \bm{\Phi}_{0:n-1},\,\bm{\Lambda}_{\text{a}}) = \prod_{n=1}^N \ist \prod_{k=1}^K p(\bm{\Phi}_{k,n} \mid \bm{\Phi}_{k,n-1},\,\bm{\Lambda}_{\text{a},k})\\[-7mm] \nn
\end{align}
where $p(\bm{\Phi}_{k,n} \mid \bm{\Phi}_{k,n-1},\,\bm{\Lambda}_{\text{a},k})$ follows a first-order Markov process with linear dynamics. In particular,
\vspace*{-2mm}
\begin{align}\label{eq:phi-markov}
  \bm{\Phi}_{k,n} 
  \;=\; 
  \bm{T}\,\bm{\Phi}_{k,n-1} + \bm{G}\,\bm{a}_k\\[-7mm] \nn
\end{align}
where $\bm{a}_k$ is a zero-mean Gaussian random vector with precision $\bm{\Lambda}_{\text{a},k}$, i.e., 
$\bm{a}\sim \mathcal{N}(\bm{0},\,\bm{\Lambda}_{\text{a},k})$, and $\bm{\Lambda}_{\text{a}} \triangleq [\bm{\Lambda}_{\text{a},1}, \dots, \bm{\Lambda}_{\text{a},K}]$. The matrices $\bm{T}$ and $\bm{G}$ are known, for instance with a constant-velocity motion model we have
\vspace*{-2mm}
\begin{align*}
  \bm{T} 
  = 
  \begin{bmatrix}
    1 & 0 & \Delta t & 0 \\
    0 & 1 & 0       & \Delta t \\
    0 & 0 & 1       & 0 \\
    0 & 0 & 0       & 1
  \end{bmatrix},
  \quad
  \bm{G} 
  = 
  \begin{bmatrix}
    \tfrac{\Delta t^2}{2} & 0 & 0 & 0 \\
    0 & \tfrac{\Delta t^2}{2} & 0 & 0 \\
    0 & 0 & \Delta t & 0 \\
    0 & 0 & 0 & \Delta t
  \end{bmatrix}\ist.\\[-7mm] \nn
\end{align*}

\subsubsection{Evolution of Existence Indicator}

We model $\xi_{k,n}$ as a discrete time birth-death process independent across $k$, i.e.,
\vspace{-2mm}
\begin{align}
    p\bigl(\xi_{k,n} \mid \xi_{k,n-1}\bigr) =
	\begin{cases}
		p_\text{s}, 	&\xi_{k,n}= 1,\ \xi_{k,n-1}=1\\
		1 - p_\text{s}, &\xi_{k,n} = 0,\ \xi_{k,n-1}=1 \\        
		p_\text{b}, 	&\xi_{k,n} = 1,\ \xi_{k,n-1}=0 \\
		1 - p_\text{b}, &\xi_{k,n}= 0,\ \xi_{k,n-1}=0 
	\end{cases}
	\label{eq:xi-markov1} \\[-7mm]\nn
\end{align}
where $p_s$, and $p_b$ are the survival and birth probability respectively.
\subsubsection{Reflectivity Model}
In modeling the distribution of $\alpha_{k,n,l}$, it is well known that an object's return strength can vary significantly from one time step to the next \cite{LepRabLeG:TAES2016}. Consequently, we assume $\alpha_{k,n,l}$ to be a priori independent across time $n$, radar $l$, and PO index $k$. These coefficients are treated as nuisance parameters and are assigned a prior \gls{pdf}
\vspace{-2mm}
\begin{align}\label{eq:priorweigths}
  p(\boldsymbol{\alpha}_{n,l}\mid\bm{\gamma}_{n,l})
  =
  \prod_{l=1}^{L}\prod_{k=1}^{K}
  \mathrm{CN}\!\bigl(\alpha_{k,n,l} \,;\, 0,\, \gamma_{k,n,l}\bigr)\\[-7mm]\nn
\end{align}
with precision $\gamma_{k,n,l}$, where $\mathrm{CN}(\bm{x};\bm{\mu},\bm{\Lambda})$ denotes the \gls{pdf} of the multivariate circular complex Gaussian distribution with mean $\bm{\mu}$ and precision $\bm{\Lambda}$ evaluated at $\bm{x}$.

\subsubsection{Prior precision}

We model the prior precision $\gamma_{k,n,l}$ as independent across the \gls{po} index $k$ and radar $l$, while evolving over time $n$ according to a Markov chain. Specifically, we assume
\vspace*{-2mm}
\begin{align}\label{eq:Markov_gamma}
    p(\gamma_{k,n,l}\mid\gamma_{k,n-1,l}) = \mathrm{Gamma}\!\left(\gamma_{k,n,l};\ \eta\,\gamma_{k,n-1,l},\,\eta\right)\\[-6mm]\nn
\end{align}
which ensures that the mean of the prior at the current time step equals the previous value $\gamma_{k,n-1,l}$. Here, $\mathrm{Gamma}(x;a,b)$ denotes the Gamma \gls{pdf} of variable $x$ with shape $a$ and rate $b$. Note that setting a high $\eta$ results in a very flat likelihood wile a low $\eta$ results in a narrow likelihood and may hence suppress current data information. 

\subsubsection{Observation Model}

At each time $n$, radar $l$ obtains a measurement vector $\bm{Z}_{n,l}$. The likelihood function is given by
\vspace{-2mm}
\begin{align}
  &p\bigl(\bm{Z}_{n,l} \mid \bm{\Phi}_n,\,\bm{\xi}_n,\,\bm{\alpha}_{n,l},\bm{\Lambda}_{z,l}\bigr)\nn\\
  &\hspace{8mm} = \mathrm{CN}\ist\Bigl(
    \bm{Z}_{n,l};
    \sum_{k=1}^{K} \xi_{k,n}\,\alpha_{k,n,l}\ist\bm{S}_l(\bm{\Phi}_{k,n}),
    \bm{\Lambda}_Z
  \Bigr)\ist. \label{eq:obs-likelihood} \\[-7mm] \nn
\end{align}
because the observation noise was assumed zero mean circular complex gaussian. 
\subsubsection{Noise variance} We assume that the noise $\tilde{\bm{W}}_l$ is independent across radars $l=1,\dots,L$, stationary with respect to time steps $n=0,\dots,N$, and is a zero mean circular complex Gaussian given the precision $\lambda_{z,l}$. It is here assumed the precision matrix is $\bm{\Lambda}_{z,l} = \lambda_{z,l}\bm{I}$, and $\lambda_{z,l}$ is unknown. We assign the prior \gls{pdf}
\vspace{-2mm}
\begin{align}
    p(\lambda_{z,l}) = \text{Gamma}(\lambda_{z,l}; \ \alpha_{z,l},\beta_{z,l})\\[-7mm]\nn
\end{align}
here $\alpha_{z,l}$, and $\beta_{z,l}$ are the shape and rate parameters.

\subsubsection{Joint Posterior pdf}

Using \eqref{eq:SST}, \eqref{eq:xi-markov1}, \eqref{eq:priorweigths}, and \eqref{eq:obs-likelihood}, and further introduce $\bm{\alpha}_{0:N,1:L} \triangleq \begin{bmatrix}
    \bm{\alpha}_{0:N,1},\hdots,\bm{\alpha}_{0:N,L}
\end{bmatrix}$ and likewise for the other radar depended parameters, the joint posterior \gls{pdf} can up to a normalization constant be written as
\vspace*{-2mm}
\begin{align}
&p\bigl(\bm{\Phi}_{0:N},\,\bm{\xi}_{0:N},\,\bm{\alpha}_{0:N,1:L},\,\bm{\Lambda}_{\text{a}},\boldsymbol{\Lambda}_{Z,1:L},\bm{\gamma}_{0:N,1:L}\,|\bm{Z}_{0:N,1:L}\bigr)\nonumber\\
  & \propto \ist p(\bm{\xi}_0)\ist\biggl(\prod_{k'=1}^{K} p(\bm{\Phi}_{k',0}) \prod_{l'=1}^{L} p(\gamma_{k',0,l'})\biggr)
  \nonumber\\
  &\quad \times
  \prod_{l = 1}^{L}\prod_{n=1}^{N}\biggl(p(\bm{\alpha}_{0,l})
    p(\bm{Z}_{n,l} \mid \bm{\Phi}_n,\,\bm{\xi}_n,\,\bm{\alpha}_{n,l},\bm{\Lambda}_{Z,l})\,
    \nonumber\\
  &\quad \times    p(\bm{\alpha}_{n,l}|\bm{\gamma}_{n,l})p(\bm{\gamma}_{n,l}|\bm{\gamma}_{n-1,l})\,
    p(\bm{\xi}_n \mid \bm{\xi}_{n-1}) \nonumber\\
  &\quad \times\prod_{k=1}^{K}
      p(\bm{\Phi}_{k,n} \rmv\mid\rmv \bm{\Phi}_{k,n-1},\,\bm{\Lambda}_{\text{a},k}) \ist p(\bm{\Lambda}_{\text{a},k})\biggr) \label{eq:joint-factorization}\\[-7mm]\nn
\end{align}
where $p(\bm{\alpha}_0)$, $p(\bm{\Phi}_{k,0})$, $p(\bm{\xi}_0)$ , and $p(\bm{\gamma}_{k,0,l})$ represent the priors at time $n=0$ and $ p(\bm{\Lambda}_{\text{a},k})$ represents the prior on the process-noise precision, and $p(\bm{\Lambda}_{Z,l})$ represents the prior on the observation noise precision.
Fig.~\ref{fig:Baysian_graph} illustrates the corresponding Bayesian network. 
Based on the joint posterior \gls{pdf} in \eqref{eq:joint-factorization}, our goal is to estimate the number of objects at each time step $\mathcal{L}_n$ by checking the number of existence indicators $\xi_{k,n}$ that has a probability of being one above a certain threshold $\delta$, and then to estimate their associated, states $\bm{\Phi}_{k,n}$ together with $\alpha_{k,n,l}$, $\gamma_{k,n,l}$, $\bm{\Lambda}_{z,l}$ and $\bm{\Lambda}_{\text{a},k}$. Since calculation of the marginal posterior for each unknown parameter based on \eqref{eq:joint-factorization} is intractable, we instead resort to approximate inference.

\section{Proposed Algorithm}\label{sec:alg_construction}
This section details the development of a direct-\gls{mot} algorithm based on \gls{vmp}.

\subsection{Mean-field VMP Approach}
Since the posterior \eqref{eq:joint-factorization} is intractable, we employ a structured mean-field approach, see e.g. \cite[Ch. 10]{bishop2007}. Specifically, we approximate \eqref{eq:joint-factorization} by a factorized surrogate \gls{pdf} given by
\vspace*{-1mm}
\begin{align}\label{eq:surogate_expression}
&q(\bm{\Phi}_{0:N},\bm{\xi}_{0:N},\bm{\alpha}_{0:N,1:L},\bm{\Lambda}_{\text{a}},\bm{\Lambda}_{Z,1:L},\bm{\gamma}_{0:N,1:L}) =\prod_{l=1}^{L}q(\bm{\Lambda}_{z,l})\nn\\[-2mm] &\hspace{5mm}\times\prod_{n=0}^N q(\bm{\alpha}_{n,l})\prod_{k=1}^{K}q(\bm{\Phi}_{n,k})q(\xi_{n,k})q(\bm{\Lambda}_{\text{a},k})q(\gamma_{k,n,l})\ist.\\[-6mm]\nn
\end{align}
The surrogate \glspl{pdf} are obtained by minimizing the Kullback--Leibler (KL) divergence between the posterior \gls{pdf} in \eqref{eq:joint-factorization} and the surrogate \gls{pdf} in \eqref{eq:surogate_expression}, which is equivalent to maximizing the \gls{elbo}. This equivalence follows from the relation
\vspace{-1mm}
\begin{align}
    \text{ELBO}(q) = \ln p - D_{\mathrm{KL}}(q\|p)\\[-7mm]\nn
\end{align}
where $p$ denotes the joint posterior, i.e., the left hand side of \eqref{eq:joint-factorization}.
Under the factorized s in \eqref{eq:surogate_expression}, the optimal choice of each factor in the KL/ELBO sense is given by \cite[Ch. 10]{bishop2007}
\vspace{-1mm}
\begin{align}\label{eq:KL_min_approach}
    \ln q^\star(\mathbf{X}_i) = \mathbb{E}_{\backslash \mathbf{X}_i}\!\left[\ln p(\mathbf{X}\mid \mathbf{D})\right] + \text{const}.\\[-6mm]\nn
\end{align}
Here, $\mathbf{X}$ and $\mathbf{D}$ are stand-ins for the variabless under consideration and the data, respectively, and $\mathbb{E}_{\backslash \mathbf{X}_i}[\cdot]$ denotes the expectation with respect to $q$ over all variables except $\mathbf{X}_i$. Assuming a factorized distribution as in \eqref{eq:joint-factorization} leads, after exponentiation, to a factorization of the surrogate. Each resulting factor can be interpreted as a message passed to the variable under consideration and is denoted by $\epsilon^{(\bm{X}_j \rightarrow \bm{X}_i)}$. Moreover, the collection of all messages incoming to a variable $\bm{X}_i$ defines its neighborhood, denoted by $\mathcal{N}_{\bm{X}_i}$. 

Using the surrogates from \eqref{eq:KL_min_approach} yields a sequential update scheme for each marginal surrogate, which is guaranteed to converge to a local optimum of the ELBO. This convergence can, however, be slow when factors are highly correlated; in such cases, it is more advantageous to optimize the joint ELBO of the correlated factors. The derivations of the individual surrogates are provided in App.~\ref{app:div_of_qPhi}--\ref{app:div_of_qLambdaz}, resulting in the approximate marginals introduced in the following.

\subsubsection{\gls{po} State Update}

Under the assumption of a normal prior $p(\bm{\Phi}_{k,0})$ for all $k$, the surrogate $q(\bm{\Phi}_{k,n})$ becomes normal if the messages from the data to the state vector,
$\epsilon^{(\bm{Z}_{n,l}\rightarrow \bm{\Phi}_{k,n})}$, are normal. However, this is generally not the case, as shown in App.~\ref{app:div_of_qPhi}. Nevertheless, motivated by reduced computational complexity and closed-form marginals, we enforce these messages to have a Gaussian functional form by fitting a Gaussian message to the true message via minimization of the $\mathrm{KL}$ divergence. The corresponding covariance is approximated by the inverse Fisher information matrix, as detailed in App.~\ref{app:div_of_data_moments}. Consequently, the final surrogate $q(\bm{\Phi}_{k,n})$ is given by the product of Gaussian messages, and the resulting approximate marginal \gls{pdf} is given by
\vspace{-1mm}
\begin{align}\label{eq:surogate_of_PHI}
    q(\bm{\Phi}_{k,n}) = \mathrm{N}\!\left(\bm{\Phi}_{k,n};\bar{\bm{\Phi}}_{k,n},\bar{\bar{\bm{\Phi}}}_{k,n}^{-1}\right).\\[-6mm]\nn
\end{align}
Here we have introduced the notation $\bar{\cdot}$ to denote the mean and $\bar{\bar{\cdot}}$ to denote the covariance. As each message is described by Gaussian \gls{pdf}, the mean and covariance of the marginal \gls{pdf} can be obtained from the messages as
\begin{align}
    \bar{\bar{\bm{\Phi}}}_{k,n}^{-1} &= \sum_{\epsilon\in\mathcal{N}(\bm{\Phi}_{k,n})}\bar{\bar{\bm{\epsilon}}}^{-1}\\
    \bar{\bm{\Phi}}_{k,n} &= \bar{\bar{\bm{\Phi}}}_{k,n}\sum_{\epsilon\in\mathcal{N}(\bm{\Phi}_{k,n})}\bar{\bar{\bm{\epsilon}}}^{-1}\bar{\bm{\epsilon}}.
\end{align}

\subsubsection{Process-noise Update}
Assuming a diagonalizing gamma prior on $\bm{\Lambda}_{\text{a},k}$, i.e., $\bm{\Lambda}_{a,k} = \diag([\lambda_{a,k,1}\ \cdots\  \lambda_{a,k,4}]^T)$ with each $\lambda_{a,k,j}$ being mutually independent and Gamma-distributed, then as shown in App.~\ref{app:div_of_qLambdaa} the surrogate is a Gamma distribution given by
\vspace{-2mm}
\begin{align}
    q(\lambda_{a,k,j}) = \text{Gamma}(\lambda_{a,k,j};\hat{a},\hat{b}_j)\\[-7mm]\nn
\end{align}
with $\hat{a}$ and $\hat{b}_j$ defined in \eqref{eq:rate_of_qLambdaa_app} and \eqref{eq:shape_of_qLambdaa_app}. The mean of $\bm{\Lambda}_{\text{a},k}$ at time $n$, can then be found as
\vspace{-1mm}
\begin{align}
    \bigg[\bar{\bm{\Lambda}}_{a,k}\bigg]_{i,j} = \begin{cases}
        \frac{\hat{a}}{\hat{b}_j}, & i=j, \\ 0, & \text{otherwise}.
    \end{cases} 
\end{align}
It is worth noting that the precision with which we estimate for the process noise, depends on how well the neighboring $\bm{\Phi}_{k,n}$ fits to the assumed kinematic model measured using the square of the $L_2$-norm as seen in \eqref{eq:V_11_app}-\eqref{eq:V_44_app}, furthermore it is upper bounded by the variance of the surrogate posterior on $\bm{\Phi}_{k,n}$. As such the uncertainty of the kinematic estimates is propagated to process noise and only complete certainty is possible if there is also zero variance on these estimates.

\subsubsection{Complex Amplitude and Existence Indicator Updates}
The complex amplitude $\bm{\alpha}_{n,l}$ and the existence $\bm{\xi}_n$ are highly correlated due to them appearing as a product in \eqref{eq:Z_constructio}, hence any decrease in existence probability can be compensated by an increase in the amplitude and vice versa. As such employing separate minimization of the KL-divergence results in slow convergence. To accelerate convergence, we optimize the \gls{elbo} jointly with respect to the factor $q(\bm{\alpha}_{n,l})$ and $q(\bm{\xi}_{k,n})$, see App.~\ref{app:div_of_alpha_and_xi}. We further assume a complex circularly-symmetric zero-mean Gaussian prior \gls{pdf} on $\bm{\alpha}_{n,l}$ and a Bernoulli prior on $\bm{\xi}_n$ arrive at the following closed-form surrogate posteriors \gls{pdf}, i.e.,
\begin{align}
    q(\bm{\alpha}_{n,l}) &= \text{CN}(\bm{\alpha}_{n,l};\bar{\bm{\alpha}}_{n,l},\bar{\bar{\bm{\alpha}}}_{n,l}^{-1})\\
    q(\xi_{k,n}) &= \text{Bern}(\xi_{k,n};\bar{\xi}_{k,n})
\end{align}
where $\text{Bern}(\xi;\bar{\xi})=\bar{\xi}^\xi \,(1-\bar{\xi})^{1-\xi}$ denotes the pmf of the Bernoulli-distribution variable $\xi\in\{0,1\}$ with mean $\bar{\xi}$.
The moments $\bar{\bm{\alpha}}_{n,l}$, $\bar{\bar{\bm{\alpha}}}_{n,l}$, and $\bar{\xi}_{k,n}$ are given in \eqref{eq:Lambda_alpha},\eqref{eq:mu_alpha}, and \eqref{eq:update_xi}, respectively.

\subsubsection{prior precision update}

Assuming a gamma prior on $\gamma_{k,n,l}$ the posterior surrogate can be calculated as shown in App.~\ref{app:div_of_qgamma} as
\begin{equation}
    q(\gamma_{k,n,l}) = \text{Gamma}(\gamma_{k,n,l};\chi_{k,n,l},\Theta_{k,n,l})
\end{equation}
with $\chi_{k,n,l}$ and $\Theta_{k,n,l}$ given in \eqref{eq:theta_def}, and \eqref{eq:Theta_def}, and mean precision given as $\bar{\gamma}_{k,n,l} = \chi_{k,n,l}/\Theta_{k,n,l}$.

\subsubsection{Observation noise update}

By again imposing a gamma prior we arrive at a closed form solution to the observation noise as shown in App.~\ref{app:div_of_qLambdaz},
\begin{equation}\label{eq:surogate_of_Lz}
    q(\lambda_{z,l}) = \text{Gamma}(\lambda_{z,l};N_s+N+1,\mathbb{W}_l)
\end{equation}
where $N_s$ is the number of collected samples and $\mathbb{W}_l$ is given in \eqref{eq:W_def}.
To summarize, the messages obtained above \eqref{eq:surogate_of_PHI}-\eqref{eq:surogate_of_Lz}, have been achieved by choosing priors carefully and restricting messages to specific functional forms. In turn this results in an message passing algorithm where all messages and marginal surrogates have known parametric forms.

After deriving update messages for all surrogates, we can construct a variety of algorithms that update the surrogates iteratively with different schedules, each corresponding to a different algorithm. Below, we introduce some approximations and define an update schedule to obtain a practical algorithm.

\subsection{Practical Implementation Aspects}

In our prior work \cite{Kitchen2025}, we have shown that the updates of $q(\bm{\Phi}_{k,n})$ do not require the full data vector $\bm{Z}_{n,l}$, but can instead be performed using only the message sent from the data\footnote{In \cite{Kitchen2025} we considered only a single object, but the conclusion generalizes to multiple objects.}, i.e., $\epsilon^{(\bm{Z}_n\rightarrow\bm{\Phi}_{k,n})}$. The information exchange then amounts to passing $\bar{\bm{\Phi}}_{k,n}$ and $\bar{\bar{\bm{\Phi}}}_{k,n}$ between radar nodes and therefore results in a low communication overhead. These compact messages enable the implementation of a redundant algorithm in which each radar node is equipped with a processing unit that runs the algorithm in parallel. Consequently, if radar nodes drop from the network, the remaining nodes can continue to operate. However, due to the updates of $q(\bm{\xi}_n)$, $q(\bm{\alpha}_{n,l})$, and $q(\bm{\gamma}_{n,l})$, this redundant approach is initially infeasible. The updates of these surrogates depend explicitly on $\bm{Z}_{n,l}$ and furthermore contain terms propagating from time $n$ to $n-1$. As a result, one would have to store $\bm{Z}_{n,l}$ for all time steps in order to update these surrogates properly, which is challenging given typical radar data rates. For this reason, we introduce the approximation that each message related to $\xi$, $\alpha$, and $\gamma$ that propagates backward in time is set to unity, and we retain only the current data vector $\bm{Z}_{N,l}$, which can be discarded at the end of each iteration.

\begin{algorithm}[t]
\caption{VMP for multi object detection and tracking}\label{alg:cap}
\begin{algorithmic}
\State \textbf{Input:} Signal vector $\bm{Z}_{N,l}$, all prior data messages $\{\epsilon_{g,k,n}\}_{n<N,\forall k,\forall l}$ and the current messges at the other sensors $\{\epsilon_{g,k,N}\}_{\forall k,l'\neq l}$, and cardinality $\hat{K}_N\leftarrow \hat{K}_{N-1}$ 

\State \textbf{Output:} Posterior marginals $q(\bm{\Phi}_{0:N})$, $q(\bm{\alpha}_{0:N,1:L})$, $q(\gamma_{1:K,0:N,1:L})$, $q(\bm{\xi}_{0:N})$, $q(\bm{\Lambda}_{k,a})$, $q(\bm{\Lambda}_{z,1:L})$ and data messages $\epsilon_{g,1:K,N,l}$
\State 
\For{$n_I^{(1)}\leftarrow 1$ to $N_I^{(1)}$}
    \State Update $q(\gamma_{k,N,l})$ using \eqref{eq:theta_def} and \eqref{eq:Theta_def}
    \State Update $q(\bm{\alpha}_{N,l})$ using \eqref{eq:Lambda_alpha} and \eqref{eq:mu_alpha}
\EndFor
\For{$k \leftarrow 1$ to $\hat{K}_N$}
\State Calculate data message $\bar{\bm{\epsilon}}_{k,g,N},\bar{\bar{\bm{\epsilon}}}_{k,g,N}$ using \eqref{eq:KL_min_message}
\State Send and receive current object messages $\{\epsilon\}_{\forall k,n,l}$ to all other radars
\For{$n_I^{(2)} \leftarrow 0 \text{ to } N_I^{(2)}$}
\For{$n \in \{0\leq n\leq N : \bar{\xi}_{k,n} > 0\}$ }
\State update $q(\bm{\Phi}_{k,n})$ using \eqref{eq:Update_of_PHI_moments}
\EndFor
\State Update $q(\bm{\Lambda}_{k,a})$ using \eqref{eq:post_for_Lambda_a} and \eqref{eq:app_def_of_V}
\EndFor
\EndFor
\State Update $q(\bm{\xi}_N)$ using \eqref{eq:update_xi}
\State For any $\bar{\xi}_{k,n}<\delta_-$ \textbf{do:} Prune track and set $\hat{K}_N \leftarrow \hat{K}_N-1$ 
\State Update $q(\bm{\Lambda}_z)$ using 
\State \textbf{Algorithm~\ref{alg:init}}: Update $q(\boldsymbol{\alpha}_N)$, $q(\bm{\xi}_N)$, $q(\boldsymbol{\Phi}_{N})$, and $\hat{K}_N$

\end{algorithmic}
\end{algorithm}

\begin{algorithm}[t]
\caption{Initialization of new objects at time $N$}\label{alg:init}
\begin{algorithmic}
\State \textbf{Input:} Signal vector $\boldsymbol{Z}_N$, current cardinality $\hat{K}_n$, current marginals $q(\bm{\Phi}_{N})$, $q(\bm{\alpha}_{N})$, $q(\bm{\xi}_{N})$, and {threshold $\delta_+$}

\State \textbf{Output:} New marginals $q(\boldsymbol{\alpha}_N)$, $q(\bm{\xi}_N)$, and $q(\boldsymbol{\Phi}_{N})$, new cardinality $\hat{K}_\text{new}$

\State
\State $\hat{K}_{\text{new}} \leftarrow \hat{K}_n$, $k = \hat{K}_{\text{new}}$, and $\bar{\xi}_{k,N}$ $\leftarrow 10$.
\For{$l\leftarrow 1$ to $L$}
\While{$\bar{\xi}_{k,N}>\delta_+$} 
\State $k \leftarrow k + 1$ 
\State Assume $\xi_{k,N}$ is 1 then:
\State $\Check{\bm{\Phi}}_{k,N}=\arg\max_{\bm{\Phi}_{k,N}} \langle\bar{\bm{\alpha}}_{N,l}|\bar{\bar{\bm{\alpha}}}_{N,l}|\bar{\bm{\alpha}}_{N,l}\rangle-\ln{|\bar{\bar{\bm{\alpha}}}_{N,l}^{-1}|}$
\State Evaluate $\bar{\xi}_{k,N}$ using \eqref{eq:update_xi} at $\Check{\bm{\Phi}}_{k,N}$
\If{$\bar{\xi}_{k,N}>\delta_+$}
\State $\hat{K}_{\text{new}} \leftarrow \hat{K}_{\text{new}} +1$

\State Update $q(\bm{\Phi}_{k,N})$ with $\{\bar{\bm{\Phi}}_{k,N},\bar{\bar{\bm{\Phi}}}_{k,N}\}$ using \eqref{eq:KL_min_message}
\State Update $q(\bm{\alpha})$, and $q({\bm{\xi}}_{N})$ using \eqref{eq:update_alpha} and \eqref{eq:update_xi}.
\State Update $q(\gamma_{k,N,l})$ using \eqref{eq:theta_def} and \eqref{eq:Theta_def}
\EndIf
\EndWhile
\EndFor

\end{algorithmic}
\end{algorithm}

A further approximation concerns the numerical optimization in \eqref{eq:MAP_estimate}. Since the true posterior \gls{pdf} is highly multimodal, we restrict the optimization in \eqref{eq:MAP_estimate} for established tracks to seven standard deviations of the prior $p(\bm{\Phi}_{k,N}|\bm{\Phi}_{k,N-1},\bm{\Lambda}_{a,k})$, reflecting physical constraints on the acceleration that can be applied to the object. This restriction ensures that the surrogate does not approximate modes different from the main lobe and effectively acts as a kinematic filter. 
Finally, as derived in this section, the updates should be run for all possible objects $K$, which becomes computationally intractable if $K=K_{\max}=N_z$ is large. We therefore run the updates only for the estimated number of true objects $\hat{K}_n$ and only for time steps in which the object is determined to be present. To detect newly appearing objects, we introduce a dedicated step that searches for new (potential) objects to track at each time step.

\subsection{Message Passing Computation Order}

Since the Bayesian network in Fig.~\ref{fig:Baysian_graph} contains loops, there is some flexibility in the order in which variational messages are computed. Different message-passing schedules may lead to different results for the surrogate \glspl{pdf}. In this work, we employ the following message-passing order. First, the set of all objects is ordered such that $1\leq k\leq \hat{K}_n$ corresponds to true objects. The algorithm is then divided into two parts.

Alg.~\ref{alg:cap} updates all surrogates for objects with nonzero probability of existence, while Alg.~\ref{alg:init} initializes new objects. Alg.~\ref{alg:cap} starts by collecting all messages related to objects with nonzero existence probability from all radar nodes in the network\footnote{Algorithms with different update strategies could also be designed, e.g., by only using data from neighboring nodes, thereby significantly reducing communication overhead.}. These messages are used to update the surrogates associated with existing objects. To ensure predictable runtime, the number of iterations of the loopy updates is restricted to fixed values $N_I^{(1)}$ and $N_I^{(2)}$, respectively. After $q(\bm{\xi}_N)$ is updated, tracks for which the probability of existence falls below a preset threshold $\delta_-$ are pruned by setting $\xi_{k,n}=0$.
Since the message from the data $\bm{Z}_{n,l}$ to the kinematic parameters $\bm{\Phi}_{k,n}$ is required for all time steps at each iteration, the moments of these messages are appended to $\bar{\bm{\epsilon}}_{g,0:N-1,l}$ and $\bar{\bar{\bm{\epsilon}}}_{g,0:N-1,l}$. These quantities can be stored locally at each radar node and form data structures of variable size $\hat{K}_n \times N \times L$. They can subsequently be reused when updating the network.

The initialization scheme described in Alg.~\ref{alg:init} is sequential. The radar nodes are therefore arbitrarily labeled $1,\dots,L$. Starting with the first radar, the first two terms of \eqref{eq:update_xi} are optimized to find the $\bm{\Phi}$ corresponding to the highest likelihood of containing a new object. Here it is assumed that $\xi_{k,N}=1$, and a prior covariance $\bm{\Sigma}_{PO}$ and amplitude precision $\gamma_{\text{init}}$ are assigned to the potential object. Subsequently, the full expression in \eqref{eq:update_xi} is optimized using the obtained $\bm{\Phi}$ to compute the mean existence probability $\bar{\xi}_{k,N}$.
If this value exceeds a preset threshold $\delta_+$, the object is added to the pool of true objects. The updated pool is then passed to the next radar node, which repeats the procedure. The final radar node $L$ transmits the resulting consensus to all radar nodes, after which the algorithm terminates.

\subsection{Complexity Analysis}

The computational complexity of our \gls{vmp} implementation is analyzed in what follows. Assuming that both algorithms are executed locally on each radar node, we begin with Alg.~\ref{alg:cap}. The first \texttt{for}-loop has a complexity of $\mathcal{O}\!\left(N_s\hat{K}_nL +(\hat{K}_n^3 +\hat{K}_n^2L)N_I^{(1)} \right)$. The calculation of the data messages has a complexity of $\mathcal{O}(\hat{K}_N^2 N_s)$. While the communication complexity associated with exchanging messages between radars is beyond the scope of this paper, the corresponding computational cost is $\mathcal{O}(\hat{K}_N)$. 

The update of $q(\bm{\Phi}_{k,0:N})$ and $q(\bm{\Lambda}_{a,k})$ is bounded by $\mathcal{O}((N+N_I^{(2)})L)$, whereas the complexity of updating $q(\bm{\xi}_N)$ is bounded by $\mathcal{O}(\hat{K}_N^3(L+1))$. The update of $q(\bm{\Lambda}_z)$ has complexity $\mathcal{O}(\hat{K}_n^2+N_s)$. Hence, the dominant term is $\hat{K}_n^3L$, as it grows with both the number of radars and the number of estimated objects.
The $\hat{K}_n^3$ complexity arises from the matrix inversion required when computing $\bar{\bm{\alpha}}$ in \eqref{eq:mu_alpha}, which reflects the correlation between signal components of different objects. Exploiting this structure could potentially reduce the computational load. Likewise, the factor $L+1$ results from incorporating information from all radar nodes when updating $q(\bm{\xi}_N)$. In practical systems, one could restrict the update to radars that are ``close'' to the estimated object position.

For Alg.~\ref{alg:init}, the complexity is dominated by the \texttt{argmax} operation used to test for the existence of new objects. This operation is upper-bounded by $\mathcal{O}(\hat{K}_n^3)$ and must be repeated for each evaluation required to find the maximum. In the implementation considered in this paper, the search is performed over a grid and therefore scales with $N_{\text{grid}}$. Consequently, in the present implementation the computational bottleneck is the number of objects, while the complexity remains otherwise linear in the remaining design parameters. If consistent runtimes are critical, one could impose a maximum number of objects $\hat{K}_{n,\max}$ or investigate more computationally efficient methods for updating $q(\bm{\alpha})$.

\section{Numerical Simulation}\label{sec:numSim}

\begin{table*}
    \centering
    \caption{Parameter settings}
    \begin{tabular}{ccccccccccc}
    \toprule
         $N_{T,R}$ & PRF          & $\mathbb{E}[\sigma_{RCS}]$          & G & Amplitude            & $R_{\text{max}}$    & $f_c$         & BW            & $T_{T_x}$       & $f_s$          & $\sigma_w^2$ \\ \midrule
        4         & 10 Hz & 0.05 $m^2$ & 30 dB & 3.9 V/m & 200 m  & 77 GHz & 96 MHz & 32 $\mu$s & 4 MHz & $10^{-6}$ \\ 
        \bottomrule 
    
    \end{tabular}    
    \label{tab:parameters}    
\end{table*}

\begin{figure}[t]
\setlength{\figurewidth}{0.75\columnwidth}
\setlength{\figureheight}{3cm}
    \centering
    \input{pgf/Crossing_fig}
    \input{pgf/Close_plot}
    \input{pgf/Handover}
    \vspace*{-3mm}
    \caption{Tracks of the considered scenarios. (a) The random crossing tracks scenario observed by four radars located on a circle at $\bm{p}_1=\begin{bmatrix}
        -100 & 160
    \end{bmatrix}^T$, $\bm{p}_2=\begin{bmatrix}
        10 & 50
    \end{bmatrix}^T$, $\bm{p}_3=\begin{bmatrix}
        -100 & -60
    \end{bmatrix}^T$, and $\bm{p}_4=\begin{bmatrix}
        -210 & 50
    \end{bmatrix}^T$. (b) Paralell track scenario with the radars placed in the same locations as shown in (a). (c) Sensor handover scenario with radars placed at $\bm{p}_i=\begin{bmatrix}
        -100+(i-1)110 & 0
    \end{bmatrix}^T$. The dashed lines denote the mean distance resulting in the return signal haveing a SNR of -20 dB.}
    \label{fig:All_Tracks_fig}
    \vspace{-3mm}
\end{figure}

\subsection{Simulation Setup}

In the simulations we consider three scenarios all with a radar system consisting of four identical $4\times4$ FMCW MIMO radars using time division multiplexing. The antennas are considered isotropic, and the transmitters have half a wavelength spacing while the receivers have wavelength spacing, and are placed such that the resulting virtual array is a uniform linear array with half wavelength spacing. The radar specifications can be found in Tab.~\ref{tab:parameters}. With the chosen settings, the classical range resolution is $1.56$~m and the maximum unambiguous range is $200$~m. At each time step $n$ the data is generated at each radar $l$ using \eqref{eq:Z_constructio}, with the RCS of the object being generated according to a Swerling-III or Swerling-$0$ model. We consider three different set of tracks as shown in Fig.~\ref{fig:All_Tracks_fig}. The initialization parameters of the priors can be found in Tab.~\ref{tab:Init_params}. For performance evaluation, we use the \gls{ospa} metric \cite{Schuhmacher2008} with a cutoff distance of $5$~m and an order of $2$. This metric is an average between cardinality error\footnote{With a cutoff distance of $5$~m if there is a mismatch between the true cardinality and the estimated cardinality this corresponds to error of $5$~m. The interested reader is referred to \cite{Schuhmacher2008} for further details on the \gls{ospa}.} and root mean square error. The \gls{ospa} is averaged over 180 simulation runs. 

\begin{table}
\caption{Algorithm initialization settings}    
    \centering
    \begin{tabular}{ccccccc}
        \toprule $p_s$ & $ p_b$ & $\delta_-$ & $\delta_+$ & $\gamma_{\text{init}}$ & $\bm{\Sigma}_{\text{PO}}$ & $\eta$ \\ \midrule
        0.92 & $10^{-3}$ & $0.01$ & 0.5 & 10 & $10\bm{I}$ & $10^{-6}/2$ \\ \bottomrule
    \end{tabular}
    \label{tab:Init_params}
\end{table}

We compare the proposed \gls{vmp} algorithm to a two-stage \gls{dtt} approach (termed SBL+BP). In the detection stage, \gls{sbl} \cite{MoeWesVenLei:Fusion2025} is applied to the data of each radar and time step, yielding object detections with location estimates. These detections are then passed to a  \gls{bp}-based \gls{mot} method \cite{MeyerProc2018,LiLeiVenTuf:TWC2022}, which performs object tracking, including track initiation, termination, and data association between detections and existing or newly formed tracks. Note that all considered scenarios study rather challenging conditions (low \gls{snr}, closely spaced objects, etc.). Additional simulations not included here show that the performance of both algorithms is almost identical in less challenging scenarios, e.g., with well separated objects and medium to high \glspl{snr}. Herein, we define the single sensor \gls{snr} for a single object located at some location $\bm{\Phi}$ as
\begin{equation}
    \text{SNR}(\bm{\Phi}) = \frac{|\alpha|^2|\bm{S}(\bm{\Phi})|^2}{\lambda_z^{-1}N_sN_TN_R}\ist.
\end{equation}

\subsubsection{Random crossing tracks}

We consider a scenario as depicted in Fig.~\ref{fig:All_Tracks_fig}a, here three objects are observed for $20$~s by four radars, the three objects are randomly started within the first $2$~s of observation time on different thirds of a circle of radius $60$~m centered at $(-100,\ist50)$~m. The tracks are evolved according to \eqref{eq:phi-markov} using a starting speed of $6$~m/s and a starting direction pointed towards the circle center. The process noise is considered white with precision matrix $\bm{\Lambda}_{\text{a}} = 6.25\bm{I}$.  The power of the radars are set such that the crossing happens close to the array gain limit for all four radars with an average \gls{snr} of around $-18$~dB. This is to test how the algorithm in low \gls{snr} scenarios on highly correlated signals.

\subsubsection{Parallel tracks}

We consider a scenario as depicted in Fig.~\ref{fig:All_Tracks_fig}b, here two objects are observed for $20$~s by four radars placed as shown in Fig.~\ref{fig:All_Tracks_fig}a. The tracks are started at $(-129, 57)$~m and $(-129, 42)$~m with an initial velocity of $\bm{v} = \begin{bmatrix}
    3 & \pm3
\end{bmatrix}^T$~m/s with the sign chosen such that the the objects are set to collide. At two seconds there is applied a constant acceleration of $\bm{a} = \begin{bmatrix}
    0 & \pm 2.71
\end{bmatrix}^T $~m/s$^2$ for one second, avoiding collision. The resulting two parallel tracks are separated by $1.5$~m, then at second $17$ $-\bm{a}$ is applied for one second to separate the tracks again. The track is constructed to be near the array gain limit of all radars, and furthermore to be observable mainly in range for two radars, and mainly in angle for the two others. The main goal is to test how the proposed \gls{vmp} algorithm performs on tracks when the objects are below the classical resolution for many time steps given initially well separated objects.

\subsubsection{Sensor handover}
We consider a scenario as depicted in Fig.~\ref{fig:All_Tracks_fig}c where a single object starts at $\bm{p} = \begin{bmatrix}
    319 & 75
\end{bmatrix}^T$~m, flies with constant velocity $\bm{v} = \begin{bmatrix}
    -8.75& 0
\end{bmatrix}^T$~m/s and is being observed for 10 minutes. The purpose of this track is to test how the track is handed over between sensors as it is only observed by no more than two sensors at a time.

\def\plotlinewidth{1pt}

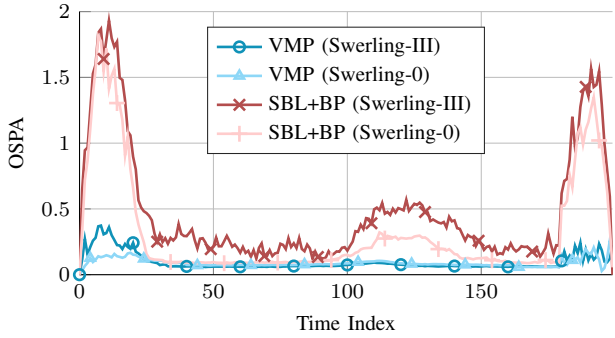
\begin{figure}[t]
    \setlength{\figurewidth}{0.8\columnwidth}
    \setlength{\figureheight}{3.5cm}
    \centering
    \begin{tikzpicture}
\begin{axis}[%
width=\figurewidth,
height=\figureheight,
at={(0\figurewidth,0\figureheight)},
scale only axis,
axis background/.style={fill=white},
axis x line*=bottom,
axis y line*=left,
xmin=0,
xmax=199,
ymin= 0,
ymax= 2,
xmajorgrids,
xminorgrids,
ymajorgrids,
yminorgrids,
axis equal=false,
xlabel={Time Index},
xlabel style={font=\footnotesize},
x tick label style={font=\footnotesize},
ylabel={OSPA},
ylabel style={font=\footnotesize},
y tick label style={font=\footnotesize},
legend style={legend cell align=left, align=left, draw=white!15!black,font=\footnotesize,anchor=north, at={(rel axis cs: 0.5,1)}, yshift=-2mm},
legend columns=1,
]

\addplot [plot3, line width=\plotlinewidth, mark=o, mark repeat=20] table [x=timeindex, y=meanVMP, col sep=comma] {pgf/CrossingTrackOSPA.csv};
\addlegendentry{VMP (Swerling-III)};
\addplot [plot1, line width=\plotlinewidth, mark=triangle, mark repeat=20,mark phase=5] table [x=timeindex, y=meanVMP, col sep=comma] {pgf/CrossingTrackOSPA_NF.csv};
\addlegendentry{VMP (Swerling-0)};
\addplot [plot4, line width=\plotlinewidth, mark=x, mark repeat=20, mark phase=10, mark size=3] table [x=timeindex, y=meanBP, col sep=comma] {pgf/CrossingTrackOSPA.csv};
\addlegendentry{SBL+BP (Swerling-III)};
\addplot [plot2, line width=\plotlinewidth, mark=+, mark repeat=20, mark phase=15, mark size=3] table [x=timeindex, y=meanSBL, col sep=comma] {pgf/CrossingTrackOSPA_NF.csv};
\addlegendentry{SBL+BP (Swerling-0)};








\end{axis}
\end{tikzpicture}
    \caption{The OSPA error for the random crossing tracks} 
    \label{fig:mean_ospa_crossing}
\end{figure}

\begin{figure}[t]
    \setlength{\figurewidth}{0.8\columnwidth}
    \setlength{\figureheight}{3.5cm}
    \centering
    \begin{tikzpicture}
\begin{axis}[%
width=\figurewidth,
height=\figureheight,
at={(0\figurewidth,0\figureheight)},
scale only axis,
axis background/.style={fill=white},
axis x line*=bottom,
axis y line*=left,
xmin=0,
xmax=199,
ymin= 0,
ymax= 2.5,
xmajorgrids,
xminorgrids,
ymajorgrids,
yminorgrids,
axis equal=false,
xlabel={Time Index},
xlabel style={font=\footnotesize},
x tick label style={font=\footnotesize},
ylabel={OSPA},
ylabel style={font=\footnotesize},
y tick label style={font=\footnotesize},
legend style={legend cell align=left, align=left, draw=white!15!black,font=\footnotesize,anchor=north, at={(rel axis cs: 0.5,1)},yshift=-2mm},
legend columns=1,
]

\addplot [plot3, line width=\plotlinewidth, mark=o, mark repeat=20, mark phase=10] table [x=timeindex, y=meanVMP, col sep=comma] {pgf/closeTrackOSPA.csv};
\addlegendentry{VMP (Swerling-III)};
\addplot [plot4, line width=\plotlinewidth, mark=x, mark repeat=20, mark phase=10, mark size=3] table [x=timeindex, y=meanBP, col sep=comma, skip coords between index={0}{1}] {pgf/closeTrackOSPA.csv};
\addlegendentry{SBL+BP (Swerling-III)};








\end{axis}
\end{tikzpicture}
    \caption{The OSPA error for the parallel tracks maneuver. The results are averaged over 180 simulation runs.}
    \label{fig:mean_ospa_close}
\end{figure}
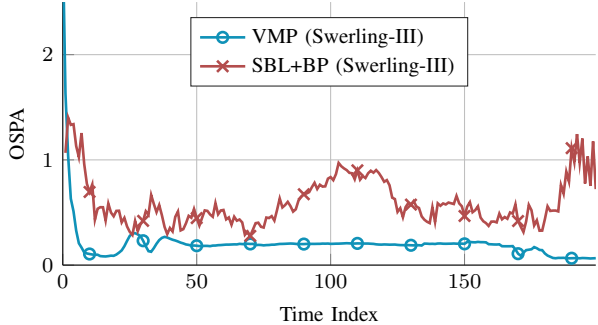

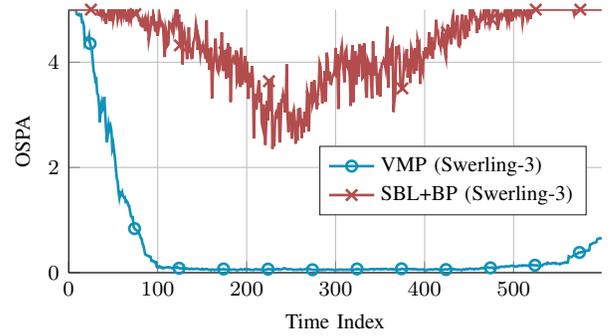
\begin{figure}[t]
    \setlength{\figurewidth}{0.8\columnwidth}
    \setlength{\figureheight}{3.5cm}
    \centering
    \begin{tikzpicture}
\begin{axis}[%
width=\figurewidth,
height=\figureheight,
at={(0\figurewidth,0\figureheight)},
scale only axis,
axis background/.style={fill=white},
axis x line*=bottom,
axis y line*=left,
xmin=0,
xmax=599,
ymin= 0,
ymax= 5,
xmajorgrids,
xminorgrids,
ymajorgrids,
yminorgrids,
axis equal=false,
xlabel={Time Index},
xlabel style={font=\footnotesize},
x tick label style={font=\footnotesize},
ylabel={OSPA},
ylabel style={font=\footnotesize},
y tick label style={font=\footnotesize},
legend style={legend cell align=left, align=left, draw=white!15!black,font=\footnotesize,anchor=east, at={(rel axis cs: 1,0.35)}, xshift=-2mm},
legend columns=1,
]

\addplot [plot3, line width=\plotlinewidth, mark=o, mark repeat=50, mark phase=25] table [x=timeindex, y=meanVMP, col sep=comma] {pgf/handoffTrackOSPA.csv};
\addlegendentry{VMP (Swerling-3)};
\addplot [plot4, line width=\plotlinewidth, mark=x, mark repeat=50, mark size=3, mark phase=25] table [x=timeindex, y=meanBP, col sep=comma, skip coords between index={0}{1}] {pgf/handoffTrackOSPA.csv};
\addlegendentry{SBL+BP (Swerling-3)};








\end{axis}
\end{tikzpicture}
    \caption{The OSPA eror for the sensor handover track. The results are averaged over 180 simulation runs..}
    \label{fig:mean_osp_handoff}
\end{figure}

\subsection{Results}

Scrutiny of the \gls{ospa} error obtained in simulations reveals that the proposed \gls{vmp} algorithm generally outperforms SBL+BP. This observation is especially apparent in the crossing scenario as reported in Fig.~\ref{fig:mean_ospa_crossing}. As the objects are in close proximity, the signals become highly correlated which results in erroneous cardinality estimates from the SBL detection algorithm, which is carried over to the BP-based tracking algorithm where the combination of low \gls{snr} and poor location estimates results in tracks being terminated early. The proposed \gls{vmp} algorithm compensates for the amplitude fluctuations more easily by directly fusing the data from all sensors in the detection step. This allows for the sensor diversity to compensate sensors which exhibit poor \gls{snr} by sensors with high \gls{snr}.

For the closely-spaced tracks shown in Fig.~\ref{fig:mean_ospa_close}, we also see an increase in the \gls{ospa} error for SBL+BP around time index $n = 100$ due to cardinality errors stemming from the two objects being close for an extended period of time leading to dropped tracks. The proposed \gls{vmp} algorithm, however, shows consistent low \gls{ospa} error with the main error stemming from detection in the first timestep and when a high acceleration is applied to the track. We observe that the average RMSE error is higher for the closely-spaced tracks than for the crossing tracks ($\sim 0.2$~m compared to $\sim 0.06$~m). This is due to the repelling effect of the function appearing in the optimization problem in \eqref{eq:MAP_estimate}. As such, the estimate for two closely-spaced objects have a tendency to ``push'' each-other apart due to their respective uncertainties. This effect is observed in the simulation results with the objects being slightly displaced along the $y$ axis for the parallel part of the tracks.

Fig.~\ref{fig:mean_osp_handoff} shows the \gls{ospa} error for the handover track. It is evident that the proposed \gls{vmp} algorithm outperforms the \gls{dtt} approach (\gls{sbl}+\gls{bp}) in low-\gls{snr} scenarios. For \gls{sbl}+\gls{bp}, the \gls{ospa} error is dominated by cardinality errors caused by low signal strength and the fact that the object is reliably observed by only a single radar at a time. In contrast, the proposed algorithm better handles the resulting amplitude fluctuations, enabling seamless handover between nodes.


An overall observation in the comparison of the two algorithms is that \gls{sbl}+\gls{bp} struggles in low-\gls{snr} scenarios (Fig.~\ref{fig:mean_ospa_crossing}--\ref{fig:mean_osp_handoff}), partly due to the large amplitude fluctuations of the Swerling-III model. The \gls{snr} can vary by up to $10$~dB between time steps, often resulting in received signal power below the detection threshold of the SBL detection stage. This effect is less pronounced in scenarios where multiple radars observe the object (Fig.~\ref{fig:mean_ospa_crossing} and Fig.~\ref{fig:mean_ospa_close}), where data fusion exploits the diversity in RCS realizations to track the object near the array limit. The proposed \gls{vmp} algorithm outperforms SBL+BP in these cases by leveraging the full detection history and directly tracking the signal evolution together with the amplitude fluctuations, thereby improving robustness to low signal strengths. As shown in Fig.~\ref{fig:mean_ospa_crossing}, under the less challenging non-fluctuating Swerling-0 model, SBL+BP performs on par with the proposed \gls{vmp} algorithm for well-separated objects.

\begin{table}
    \centering
    \caption{Runtime of the Crossing tracks}
    \begin{tabular}{lcccc}
       \toprule Algorithm & Mean [s] & Standard deviation [s] & Min [s] &  Max [s] \\ \midrule 
         VMP & 1.29 & 0.37 & 0.29 & 3.77  \\
         SBL + BP & 1.59 & 0.99 & 0.008 & 17.1 \\
         \qquad SBL & 1.53 & 0.97 & 0.006 & 17.1 \\
         \qquad BP & 0.05 & 0.03 & 0.002 & 0.18 \\ \bottomrule
    \end{tabular}
    \label{tab:runtimes}
\end{table}
We examine the computational complexity by measuring the runtime in our Monte Carlo implementation. Tab.~\ref{tab:runtimes} reports the mean average runtime, standard deviation, minimum, and maximum values over 16 runs of the crossing tracks scenario which corresponds to 3200 algorithm calls. The experiments are conducted on an Apple M1 Pro chip. The results show that the runtime of the SBL+BP method is dominated by the SBL algorithm, which exhibits higher mean runtime compared to the proposed VMP algorithm. In contrast, the \gls{bp}-based \gls{mot} algorithm contributes only a small fraction to the overall runtime. Although the combined runtime of SBL+BP is higher in the mean and maximum categories, the results indicate that the proposed \gls{vmp} algorithm does not increase computational complexity and remains suitable for potential real-time implementations.

\section{Conclusion}

We proposed a direct-\gls{mot} method based on \gls{vmp} that operates directly on raw MIMO radar signals and avoids the classical detect-then-track processing pipeline. By modeling the superimposed radar signal and performing inference directly at the signal level, the proposed approach jointly estimates object existence, kinematic states, and nuisance parameters such as object reflectivities and noise variance within a unified probabilistic framework. The method further enables coherent data fusion across multiple radar nodes and accounts for the reliability of individual radar-object links through a hierarchical Bernoulli-Gamma model. The resulting \gls{vmp} algorithm yields computationally efficient message updates under a mean-field approximation while explicitly capturing correlations in the radar measurements caused by closely spaced objects. This direct signal-level inference allows the method to preserve weak signal components that may be discarded in conventional pipelines relying on intermediate detections.

Simulation results in challenging low-\gls{snr} scenarios demonstrate that the proposed approach achieves significantly lower \gls{ospa} errors than a conventional \gls{dtt} pipeline based on super-resolution \gls{sbl} estimation followed by \gls{bp} tracking. The improvements are particularly pronounced when multiple objects are closely spaced and produce strongly correlated measurements across sensors.

Promising directions for future research including extensions of the proposed framework toward hybrid inference approaches that combine model-based Bayesian inference with data-driven components \cite{GreSteSch:NeuRIPS2017,VenLeiTerWit:JSP2023,WeiLiaMey:TSP2026}. 

\appendices
\section{Derivation of $q(\bm{\Phi}_{k,n})$}\label{app:div_of_qPhi}
Expressing \eqref{eq:KL_min_approach} using \eqref{eq:joint-factorization} and \eqref{eq:surogate_expression} we arrive at
\vspace{-2mm}
\begin{multline}\label{eq:sorrogate_of_Phi_Bishop}
    \ln{q(\bm{\Phi}_{k,n})} = \sum_{l=1}^L\mathbb{E}_{\backslash \bm{\Phi}_{k,n}}[\ln p(\boldsymbol{Z}_{n,l}|\boldsymbol{\Phi}_{k,n},\boldsymbol{\xi}_{n},\boldsymbol{\alpha}_{n,l})]\\+ \mathbb{E}_{\backslash \bm{\Phi}_{k,n}}[\ln p(\bm{\Phi}_{k,n}|\bm{\Phi}_{k,n-1},\bm{\Lambda}_{k,a})] \\+ \mathbb{E}_{\backslash \bm{\Phi}_{k,n}}[\ln p(\bm{\Phi}_{k,n+1}|\bm{\Phi}_{k,n},\bm{\Lambda}_{k,a})] + \text{constant}.
\end{multline}
The first term $\ln\epsilon^{(\bm{Z}_{n,l}\rightarrow\bm{\Phi}_{k,n})}$ does not have a closed form expression as shown in previous work \cite{Kitchen2025}. 
To obtain closed form solutions for all marginals, we restrict $\epsilon^{(\bm{Z}_{n,l}\rightarrow\bm{\Phi}_{k,n})}$ to be a Gaussian which minimizes the KL divergence w.r.t. the true message as the Gaussian is fully defined by its mean and covariance this is written as
\vspace*{-1mm}
\begin{align}\label{eq:KL_min_message}
    \{\bar{\bm{\epsilon}}_{g,k,n,l},\bar{\bar{\bm{\epsilon}}}_{g,k,n,l}\} = \underset{\bar{\bm{\epsilon}}_g,\bar{\bar{\bm{\epsilon}}}_g}{\arg\min} \, \mathcal{D}_{KL}({\epsilon}_g||{\epsilon}^{(\bm{Z}_{n,l} \rightarrow \bm{\Phi}_{k,n})}).
\end{align}
Here ${\epsilon}_g$ is the Gaussian message, $\bar{\cdot}$ denotes the mean and $\bar{\bar{\cdot}}$ the covariance matrix. It is known that a minimization of the KL divergence between a gaussian and an arbitrary distribution is the same as matching the first two moments of the Gaussian to the distribution \cite{Gerhard2016}. We here find the first moment as the maximum posterior probability of $\epsilon^{(\bm{Z}_{n,l}\rightarrow\bm{\Phi}_{k,n})}$ and approximate the covariance $\bar{\bar{\bm{\epsilon}}}_g$ using the fisher information matrix \eqref{eq:Fisher_information_matrix} as outlined in App.~\ref{app:div_of_data_moments}.

The second and third terms in \eqref{eq:sorrogate_of_Phi_Bishop} yield the following messages
\vspace{-2mm}
\begin{align} 
    &\epsilon^{(\bm{\Lambda}_a,\bm{\Phi}_{k,n-1}\rightarrow\bm{\Phi}_{k,n})}\nn\\ &\hspace{5mm}=  \mathrm{N}\left(\bm{\Phi}_{k,n};\bm{T}\bar{\bm{\Phi}}_{k,n-1},\bm{G}^{-T}\bar{\bm{\Lambda}}_{k,a}\bm{G}^{-1}\right)  \\
    &\epsilon^{(\bm{\Lambda}_a,\bm{\Phi}_{k,n+1}\rightarrow\bm{\Phi}_{k,n})} \nn\\
    &\hspace{5mm}= \mathrm{N}\left(\bm{\Phi}_{k,n};\bm{T}^{-1}\bar{\bm{\Phi}}_{n+1}^{(k)},\bm{T}^{T}\bm{G}^{-T}\bar{\bm{\Lambda}}_{k,a}\bm{G}^{-1}\bm{T}\right).\\[-7mm]\nn
\end{align}
Notice that all messages to update $q(\bm{\Phi}_{k,n})$ are now Gaussian and hence $q(\bm{\Phi}_{k,n})$ is a product of Gaussian which is also itself Gaussian with the following mean and precision matrix
\begin{equation}\label{eq:Update_of_PHI_moments}
    \bar{\bar{\bm{\Phi}}}_{k,n}^{-1} =\hspace{-3mm} \sum_{\epsilon\in\mathcal{N}(\bm{\Phi}_{k,n})}\hspace{-3mm} \bar{\bar{\bm{\epsilon}}}^{-1}, \phantom{mm} \bar{\bm{\Phi}}_{k,n} = \bar{\bar{\bm{\Phi}}}_{k,n} \hspace{-3mm}\sum_{\epsilon\in\mathcal{N}(\bm{\Phi}_{k,n})}\hspace{-3mm}\bar{\bar{\bm{\epsilon}}}_{k,n}^{-1}\bar{\bm{\epsilon}}
\end{equation}

\section{Derivation of $q(\bm{\Lambda}_{a,k})$}\label{app:div_of_qLambdaa}
Expressing \eqref{eq:KL_min_approach} using \eqref{eq:joint-factorization} and \eqref{eq:surogate_expression} we get
\begin{multline}\label{eq:surrogate_of_Lambda_a}
    \ln{q(\bm{\Lambda}_{a,k})} = \text{const} + \ln{p(\bm{\Lambda}_{a,k})} \\+ \sum_{n=0}^{N}\mathbb{E}_{\backslash \bm{\Lambda}_{a,k}}[\ln{p(\bm{\Phi}_{k,n}|\bm{\Phi}_{k,n-1},\bm{\Lambda}_{a,k})}]\mathds{1}_{n\neq 0}.
\end{multline} 
We write out the the second term of \eqref{eq:surrogate_of_Lambda_a} as
\begin{multline}\label{eq:app_message_for_Lambda_a}
    e^{\sum_{n=1}^{N}\mathbb{E}_{\backslash\bm{\Lambda}_{a,k}}[\ln p(\bm{\Phi}_{k,n}|\bm{\Phi}_{k,n-1},\bm{\Lambda}_{a,k})]}\\= e^{-\frac{1}{2}\sum_{n=1}^{N}Tr(\bm{\Lambda}_{a,k}\mathbb{V}_{k,n,n-1})}|\bm{\Lambda}_{a,k}|^{N/2}
\end{multline}
\begin{multline}\label{eq:app_def_of_V}    
    \mathbb{V}_{k,n,n-1} = \mathbb{E}_{\bm{\Phi}_{k,n},\bm{\Phi}_{k,n-1}}[\bm{G}^{-1}|\bm{\Phi}_{k,n}-\bm{T}\bm{\Phi}_{k,n-1}\rangle\\\times \langle\bm{\Phi}_{k,n}-\bm{T}\bm{\Phi}_{k,n-1}|\bm{G}^{-T}].
\end{multline}
Here $|\cdot\rangle\langle\cdot|$ is the bra-ket notation for outer products. By imposing a factorized gamma \gls{pdf} for prior \gls{pdf}, i.e., $p([\bm{\Lambda}_{a,k}]_{j,j}) = \prod_{j=1}^4 p(\lambda_{a,k,j}) = \prod_{j=1}^4\text{Ga}(\lambda_{a,k,j}|a,\ist b)$ with shape parameter $a=\zeta/2$ and scale parameter $b=\chi/2$. Writing the trace as
\begin{equation}
     Tr(\bm{\Lambda}_{a,k}\mathbb{V}_{k,n,n-1}) = \sum_{j=1}^4 \lambda_{a,k,j}[\mathbb{V}_{k,n,n-1}]_{j,j}
\end{equation}
resulting in \eqref{eq:app_message_for_Lambda_a} becoming completely separable in $\lambda_{a,k,j}$ as $f(\lambda_{a,k,1},\hdots,\lambda_{a,k,4}) = \prod_j f(\lambda_{a,k,j})$ with,
\begin{equation}
     f(\lambda_{a,k,j}) = (\lambda_{a,k,j})^{N/2}e^{-\frac{1}{2}\lambda_{a,k,j}\sum_{n=1}^N[\mathbb{V}_{k,n,n-1}]_{j,j}}.
\end{equation}
Noting that this has the functional form of a gamma distribution, taking the expectation in \eqref{eq:app_def_of_V} yields the following expressions for $[\mathbb{V}_{k,n,n-1}]_{j,j}$
\begin{multline}\label{eq:V_11_app}
    [\mathbb{V}_{k,n,n-1}]_{1,1} = \frac{4}{\Delta t^4}[|\bar{x}_{k,n}-\bar{x}_{k,n-1}|^2 +\sigma_{x,k,n}^2 \\+ \sigma_{x,k,n-1}^2+\Delta t^2 \sigma_{v_x,k,n-1}^2-2\Delta t\,\sigma_{x,v_x,k,n-1}]
\end{multline}
\begin{multline}\label{eq:V_22_app}
    [\mathbb{V}_{k,n,n-1}]_{2,2} = \frac{4}{\Delta t^4}[|\bar{y}_{k,n}-\bar{y}_{k,n-1}|^2 +\sigma_{y,k,n}^2 \\+ \sigma_{y,k,n-1}^2 +\Delta t^2 \sigma_{v_y,k,n-1}^2-2\Delta t\,\sigma_{y,v_y,k,n-1}]
\end{multline}
\begin{equation}\label{eq:V_33_app}
    [\mathbb{V}_{k,n,n-1}]_{3,3} =\hspace{-0.1cm} \frac{1}{\Delta t^2}[|\bar{v}_{x,k,n}-\bar{v}_{x,k,n-1}|^2 + \sigma_{v_x,k,n}^2 + \sigma_{v_x,k,n-1}^2]
\end{equation}
\begin{equation}\label{eq:V_44_app}
    [\mathbb{V}_{k,n,n-1}]_{4,4} =\hspace{-0.1cm} \frac{1}{\Delta t^2}[|\bar{v}_{y,k,n}-\bar{v}_{y,k,n-1}|^2 + \sigma_{v_y,k,n}^2 + \sigma_{v_y,k,n-1}^2].
\end{equation}
The surrogate \gls{pdf} is
\vspace{-1mm}
\begin{align}\label{eq:post_for_Lambda_a}
    q(\lambda_{a,k,j}) = \text{Ga}\left(\lambda_{a,k,j};\hat{a},\ist\hat{b}_j\right)\\[-7mm]\nn
\end{align}
with the rate and shape parameters
\begin{align}
    \hat{a} &= (N+\zeta)/2\label{eq:rate_of_qLambdaa_app}\\
    \hat{b}_j &= \frac{1}{2}(\chi + \sum_{n=1}^{N}[\mathbb{V}_{k,n,n-1}]_{j,j}).
    \label{eq:shape_of_qLambdaa_app}
\end{align}

\section{Derivation of $q(\bm{\alpha}_{n,l})$ and $q(\xi_{k,n})$}\label{app:div_of_alpha_and_xi}

As $\bm{\alpha}_{n,l}$ and $\bm{\xi}_{n}$ is highly correlated we seek to maximize the joint ELBO, as this was found to yield computational faster results \cite{Moederl2024,Badiu2017}
\begin{align}
    \text{ELBO}(q) &= \mathbb{E} \big[\ln p\bigl(\bm{\Phi}_{0:N},\,\bm{\xi}_{0:N},\,\bm{\alpha}_{0:N,1:L},\,\bm{\Lambda}_a,\nn\\  &\hspace{10mm}\boldsymbol{\Lambda}_{Z,1:L},\bm{\gamma}_{0:N,1:L}\,|\bm{Z}_{0:N,1:L}\bigr)\big] + H(q)\label{eq:ELBO_main_article}
\end{align}
where $H(\cdot)$ is the entropy, and the expectation is with respect to the surrogate \gls{pdf} in \eqref{eq:surogate_expression}. Further expanding
\begin{multline}\label{eq:app_elbo}
    \text{ELBO}(q) = \ln{G(\bar{\bm{\xi}}_n)} +\sum_{k=1}^K H(q(\xi_{k,n})) + \bar{\xi}_{k,n} g(\bar{\xi}_{k,n-1}) \\ - D_{KL}(q(\bm{\alpha}_{n,l})||t) + C
\end{multline}
with
\begin{equation}\label{eq:g_def}
    g(\bar{\xi}_{k,n-1}) = \bar{\xi}_{k,n-1}(\text{logit}(p_s)-\text{logit}(p_b)) + \text{logit}(p_b)
\end{equation}
$\text{logit}(x) = \ln{\frac{x}{1-x}}$, and $G(\bar{\bm{\xi}}_n)$ as
\begin{equation}\label{eq:app_integral_for_G}
    G(\bar{\bm{\xi}}_n) = \int_{\bm{\alpha}_{n,l}} e^{\mathbb{E}_{\backslash \boldsymbol{\alpha}_{n,l}}[\ln{p(\bm{Z}_{n,l}|\bm{\Phi}_n,\bm{\xi}_n,\bm{\alpha}_{n,l})}]+ \ln{p(\bm{\alpha}_{n,l})}} d\bm{\alpha}_{n,l}.
\end{equation}
is the normalization of the probability distribution
\begin{equation}\label{eq:app_t_def}
    t(\bm{\alpha}_{n,l};\bar{\bm{\xi}}_n) = \frac{e^{\mathbb{E}_{\backslash \boldsymbol{\alpha}_{n,l}}[\ln{p(\bm{Z}_{n,l}|\bm{\Phi}_n,\bm{\xi}_n,\bm{\alpha}_{n,l})}]+ \ln{p(\bm{\alpha}_{n,l})}}}{G(\bar{\bm{\xi}})}.
\end{equation}
The maximum ELBO, results from letting $q(\bm{\alpha}_{n,l}) = t(\bm{\alpha}_{n,l};\bar{\bm{\xi}}_{n,l})$. The expectation in \eqref{eq:app_t_def} is of  the form
\begin{multline}\label{eq:app_Loglike_iN_Terms_of_alpha}
    \mathbb{E}_{\backslash \bm{\alpha}_{n,l}}[\ln{p(\bm{Z}_{n,l}|\bm{\Phi}_n,\bm{\alpha}_{n,l},\bm{\xi}_n)}] =-\langle \bm{Z}_{n,l}|\bar{\bm{\Lambda}}_Z|\bm{Z}_{n,l}\rangle \\-\langle\bm{\alpha}_{n,l}|\bm{M}_n\odot\mathbb{E}_{\bm{\Phi}}[\langle\bm{S}_{n,l}^T|\bar{\bm{\Lambda}}_Z|\bm{S}_{n,l}^T\rangle]|\bm{\alpha}_{n,l}\rangle \\+ \text{Re}\{\langle\bm{Z}_{n,l}|\bar{\bm{\Lambda}}_Z|\mathbb{E}_{\bm{\Phi}}[\bm{S}_{n,l}^T]\bar{\bm{\xi}}\rangle|\bm{\alpha}_{n,l}\rangle\}+f(\bar{\bm{\Lambda}}_Z).
\end{multline}
Here $\odot$ is the element wise multiplication, $\langle\cdot|\cdot\rangle$ is the bra-ket notation for inner products, and the matrices $\bm{M}_n$, $\bm{S}_{n,l}^T$, and $\bar{\bm{\xi}}$ are given as 
$\bm{S}_{n,l}^T = [\bm{S}_l(\bm{\Phi}_{1,n}) \ist\cdots\ist \bm{S}_l(\bm{\Phi}_{k,n})]$, $\bar{\bm{\xi}} = \text{diag}([\bar{\xi}_{1,n}\ist\cdots\ist \bar{\xi}_{k,n}]^T)$, and
\begin{align}
    \bm{M}_n &= \begin{bmatrix}
        \bar{\xi}_{1,n} & \bar{\xi}_{2,n}\bar{\xi}_{1,n} & \hdots & \bar{\xi}_{k,n}\bar{\xi}_{1,n} \\
        \bar{\xi}_{1,n}\bar{\xi}_{2,n} & \bar{\xi}_{2,n} & \hdots & \bar{\xi}_{k,n}\bar{\xi}_{2,n} \\
        \vdots & \vdots & \ddots & \vdots \\
        \bar{\xi}_{1,n}\bar{\xi}_{k,n} & \bar{\xi}_{2,n}\bar{\xi}_{k,n} & \hdots &\bar{\xi}_{k,n}
    \end{bmatrix}.
\end{align}
We then recognize $t(\bm{\alpha}_{n,l};\bm{\xi}_n)$ as a circular complex Gaussian with precision and mean given by
\begin{equation}\label{eq:Lambda_alpha}
    \bar{\bar{\bm{\alpha}}}_{n,l}^{-1} = \bm{M}_n\odot\mathbb{E}[\langle\bm{S}_{n,l}^T|\bar{\bm{\Lambda}}_z|\bm{S}_{n,l}^T\rangle] + \bm{\gamma}_{n,l}
\end{equation}
\begin{equation}\label{eq:mu_alpha}
    \bar{\bm{\alpha}}_{n,l} = \bar{\bar{\bm{\alpha}}}_{n,l}\langle\mathbb{E}[\mathbf{S}_{n,l}]\bar{\bm{\xi}}_n|\bar{\bm{\Lambda}}_z|\bm{Z}_{n,l}\rangle.
\end{equation}
The optimal distribution for $q(\bm{\alpha}_{n,l})$ is then
\begin{align}
        q(\bm{\alpha}_{n,l}) = \mathrm{CN}(\bm{\alpha}_{n,l}; \bar{\bm{\alpha}}_{n,l},\bar{\bar{\bm{\alpha}}}_{n,l})\label{eq:update_alpha}\\[-6mm].\nn
\end{align}
The optimal $q(\bm{\xi}_{n})$ is a Bernoulli distribution with mean $\bar{\bm{\xi}}_n$ found by direct optimization of \eqref{eq:app_elbo} 
as 
\begin{multline}\label{eq:update_xi}
    \bar{\bm{\xi}}_n = \underset{\bar{\bm{\xi}}}{\arg\max}\, \sum_{l=1}^{l}\,\langle\bar{\bm{\alpha}}_{n,l}|\bar{\bar{\bm{\alpha}}}_{n,l}^{-1}|\bar{\bm{\alpha}}_{n,l}\rangle-\ln{|\bm{\Lambda}_{\bm{\alpha},n,l}|}\\+\sum_{k=1}^{K} H(q(\xi_{k,n}))+\bar{\xi}_{k,n}g(\bar{\xi}_{k,n-1}).
\end{multline}

\section{Derivation of $q(\gamma_{k,n,l})$}\label{app:div_of_qgamma}
As shown in \eqref{eq:priorweigths} the precisions are assumed to be independent across objects, i.e., $\bm{\gamma}_{n,l} = \text{diag}(\begin{bmatrix}
    \gamma_{1,n,l} & \gamma_{2,n,l} & \hdots &  \gamma_{K,n,l}
\end{bmatrix}$. Taking the expectation in \eqref{eq:KL_min_approach} yields
\begin{multline}\label{eq:messages_for_gamma}
    q(\gamma_{k,n,l}) = \mathbb{E}_{\backslash \gamma_{k,n,l}}[\ln p(\bm{\alpha}_{n,l}|\bm{\gamma}_{n,l})] \\+ \mathbb{E}_{\backslash \gamma_{k,n,l}}[\ln p(\gamma_{k,n,l}|\gamma_{k,n-1,l})] + \text{const.}
\end{multline}
Where the message $\epsilon^{(\gamma_{k,n+1,l}\rightarrow\gamma_{k,n,l})}$ has been omitted as discussed in Sec.~\ref{sec:system_model}. Using \eqref{eq:Markov_gamma} the first term in \eqref{eq:messages_for_gamma} reads
\begin{multline}\label{eq:message_1_for_gamma}
    \mathbb{E}_{\backslash \gamma_{k,n,l}}[\ln p(\bm{\alpha}_{n,l}|\bm{\gamma}_{n,l})] \propto - \gamma_{n,k,l}(|\alpha_{n,k,l}|^2 + \sigma_{\alpha,n,k,l}^2) \\+ \ln{\gamma_{n,k,l}} + \text{const}
\end{multline}
The second term of \eqref{eq:messages_for_gamma} reads
\begin{multline}\label{eq:message_2_for_gamma}
    \mathbb{E}_{\backslash\gamma_{k,n,l}}[p(\gamma_{k,n,l}|\gamma_{k,n-1,l})] = (\eta\,\bar{\gamma}_{k,n-1,l}-1)\ln\gamma_{k,n,l}\\-\eta\,\gamma_{k,n,l} +\text{const.}
\end{multline}
Using \eqref{eq:message_1_for_gamma} and \eqref{eq:message_2_for_gamma} we see that the marginal surrogate is a gamma distribution
\begin{equation}
    q(\gamma_{n,k,l}) = \text{Gamma}(\chi_{k,n,l},\Theta_{k,n,l})
\end{equation}
with rate and shape parameters given as
\begin{equation}\label{eq:theta_def}
    \chi_{k,n,l} = \eta\,\bar{\gamma}_{k,n-1,l} + 1
\end{equation}
\begin{equation}\label{eq:Theta_def}
    \Theta_{k,n,l} = \eta\ + |\alpha_{k,n,l}|^2 +\sigma_{\alpha,k,n,l}^2.
\end{equation}

\section{Derivation of $q(\bm{\Lambda}_{z,l})$}\label{app:div_of_qLambdaz}
As stated in Sec.~\ref{sec:system_model}, we assume $\bm{\Lambda}_{z,l} = \lambda_{z,l}\bm{I}$, using this and writing out \eqref{eq:KL_min_approach} we arrive at
\begin{multline}
    \ln q(\lambda_{z,l}) = \sum_{n=0}^{N}\mathbb{E}_{\backslash\lambda_{z,l}}[\ln p(\bm{Z}_{n,l}|\bm{\Phi}_{n},\bm{\xi}_n,\bm{\alpha}_{n,l},\lambda_{z,l}\bm{I})] \\+ \mathbb{E}_{\backslash\lambda_{z,l}}[p(\lambda_{z,l})]+ \text{const.}
\end{multline}
For the first term we get
\begin{multline}\label{eq:app_calc_of_message_Z_to_Lambdaz}
    \sum_{n=0}^{N}\mathbb{E}_{\backslash\lambda_{z,l}}[\ln p(\bm{Z}_{n,l}|\bm{\Phi}_{n},\bm{\xi}_n,\bm{\alpha}_{n,l},\lambda_{z,l}\bm{I})] = -\lambda_{z,l}\mathbb{W}_l \\+ (N+N_s+1)\ln{\lambda_{z,l}}.
\end{multline}
with
\begin{multline}\label{eq:W_def}
    \mathbb{W}_l = \sum_{n=0}^N \langle\bm{Z}_{n,l}|\bm{Z}_{n,l}\rangle + \langle\bar{\bm{\alpha}}_{n,l}|\bm{M}\odot\mathbb{E}[\langle\bm{S}_{n,l}|\bm{S}_{n,l}\rangle]|\bar{\bm{\alpha}}_{n,l}\rangle \\+Tr(\bar{\bar{\bm{\alpha}}}(\bm{M}\odot\mathbb{E}[\langle\mathbf{S}_{n,l}|\mathbf{S}_{n,l}\rangle]))- 2Re\{\langle\mathbf{Z}_{n,l}|\bar{\bm{S}}_{n,l}\bar{\bm{\xi}}_{n}\bar{\bm{\alpha}}_{n,l}\rangle\}.
\end{multline}
The functional form in \eqref{eq:app_calc_of_message_Z_to_Lambdaz} is that of a gamma distribution. Combining with the prior we get
\begin{equation}
    q(\lambda_{n,l}) = \text{Gamma}(\lambda_{n,l};N_s+N+1+\alpha_{z,l},\mathbb{W}_l+\beta_{z,l}).
\end{equation}
    
\section{Derivation of $\bar{\bm{\epsilon}}_g$, and $\bar{\bar{\bm{\epsilon}}}_g$}\label{app:div_of_data_moments}
Starting from \eqref{eq:KL_min_message} we note it is the KL divergence between a Gaussian and an unknown distribution. This is achieved by matching the moments of the gaussian and $\epsilon^{(\bm{Z}_{n,l}\rightarrow\bm{\Phi}_{k,n})}$ \cite{Gerhard2016}. To this end, we choose to use the MAP estimate as the estimate of the mean, i.e.
\begin{equation}\label{eq:MAP_estimate}
    \bar{\bm{\epsilon}}_{g,k,n,l} = \underset{\bar{\bm{\Phi}}_{k,n}}{\text{argmax}} \,\,\mathbb{E}_{\backslash\bm{\Phi}_{k,n}}[\ln p(\bm{Z}_{n,l}|\bm{\alpha}_{n,l},\bm{\xi}_n,\bm{\Phi}_{n},\bm{\Lambda}_{z,l})].
\end{equation}
The expectation yields
\begin{align}\label{eq:KL_divergence}
    &\mathbb{E}_{\backslash\bm{\Phi}_{k,n}}[\ln p(\bm{Z}_{n,l}|\bm{\alpha}_{n,l},\bm{\xi}_n,\bm{\Phi}_{n},\bm{\Lambda}_{z,l})] = \text{Const} - 2\text{Re}\Big\{\big\langle\bm{Z}_n  \nn \\
    &\hspace{3mm} - \sum_{k'\neq k}\bar{\alpha}_{k',n,l}\bar{\xi}_{k',n}\bm{S}_l(\bar{\bm{\Phi}}_{k',n})|\bar{\bm{\Lambda}}_Z|\bar{\alpha}_{k,n,l}\bar{\xi}_{k,n}\bm{S}_l(\bar{\bm{\epsilon}}_{g})\big\rangle\Big\} \nn \\
    &\hspace{3mm}  + (|\bar{\alpha}_{k,n,l}|^2 + \bar{\bar{\alpha}}_{k,n,l})\bar{\xi}_{k,n}
     \bigg[ 
    \langle \bm{S}_l(\bar{\bm{\epsilon}}_{g}) |\bar{\bm{\Lambda}}_Z|\bm{S}_l(\bar{\bm{\epsilon}}_{g})\rangle \nn  \\
    &\hspace{3mm} + \text{Tr}\Big(\bar{\bar{\bm{\epsilon}}}_{g}\langle\nabla\bm{S}_l(\bm{\Phi}_{k,n})\Big|_{\bar{\bm{\epsilon}}_{g}}|\bar{\bm{\Lambda}}_Z|\nabla\bm{S}_l(\bm{\Phi}_{k,n})\Big|_{\bar{\bm{\epsilon}}_{g}}\rangle\Big)\bigg]
    .
\end{align}
We approximate the expectation of $\bm{S}(\bm{\Phi}_{k,n})$ using the delta method:
\begin{equation}
    \mathbb{E}_{\bm{\Phi}_{k,n}}[\bm{S}_l(\bm{\Phi}_{k,n})]\approx \bm{S}_l(\bar{\bm{\Phi}}_{k,n})
\end{equation}
\begin{align}
    &\mathbb{E}_{\bm{\Phi}_{k,n}}[\langle \bm{S}_l(\bm{\Phi}_{k,n}) |\bar{\bm{\Lambda}}_Z|\bm{S}_l(\bm{\Phi}_{k,n})\rangle] \approx \langle \bm{S}_l(\bar{\bm{\Phi}}_{k,n}) |\bar{\bm{\Lambda}}_Z|\bm{S}_l(\bar{\bm{\Phi}}_{k,n})\rangle \nn\\
    &\hspace*{4mm} + \text{Tr}\Big(\bar{\bar{\bm{\Phi}}}_{k,n}\langle\nabla\bm{S}_l(\bm{\Phi}_{k,n})\Big|_{\bar{\bm{\Phi}}_{k,n}}|\bar{\bm{\Lambda}}_Z|\nabla\bm{S}_l(\bm{\Phi}_{k,n})\Big|_{\bar{\bm{\Phi}}_{k,n}}\rangle\Big)
\end{align}
with $\nabla$ denoting the gradient with respect to $\bm{\Phi}_{k,n}$. 
To estimate the covariance matrix, $\bar{\bar{\bm{\epsilon}}}_g$ we use the Fisher information matrix as it is a lower bound on the variance i.e., $\bm{\Sigma}_\theta\geq \mathcal{I}^{-1}(\theta)$. 

We start by considering
\begin{multline}
\ln\epsilon(\bm{\Phi}_{k,n}) \propto \\{-\mathbb{E}_{\backslash\bm{\Phi}_{k,n}}[\langle\tilde{\mathbf{Z}}_{n,l}-\tilde{\bm{S}}_{k,n,l}({\bm{\Phi}}_{k,n})|{\bm{\Lambda}}_{z,l}|\tilde{\mathbf{Z}}_{n,l}-\tilde{\bm{S}}_{k,n,l}({\bm{\Phi}}_{k,n})\rangle]}
\end{multline}
with $\tilde{\bm{Z}}_{n,l} = \bm{Z}_{n,l}-\sum_{k'\neq k}\xi_{k',n}\alpha_{k',n,l}\bm{S}_{l}({\bm{\Phi}}_{k',n})$, and $\tilde{\bm{S}}_{k,n,l} =\xi_{k,n}\alpha_{k,n,l}\bm{S}_l(\bm{\Phi}_{k,n})$ further using $\tilde{\bm{Z}}_{n,l}\sim N^C(\bm{\mu}(\bm{\Phi}_{k,n}),\bm{\Lambda}_z)$ we have that 
\begin{equation}
    \mathcal{I}_l(\bm{\Phi}_{k,n}) = \mathbb{E}_{\tilde{\bm{Z}}_{n,l}}[\nabla\nabla^T\ln\epsilon(\bm{\Phi}_{k,n})|\bm{\Phi}_{k,n}]
\end{equation}
carrying out the differentiation we get
\begin{multline}
    \nabla\nabla^T\ln\epsilon(\bm{\Phi}_{k,n}) = -2Re\bigg\{\xi_{k,n}^2|\alpha_{k,n,l}|^2\bigg[\\ \langle\nabla\bm{S}_l(\bm{\Phi}_{k,n})|\bm{\Lambda}_{z,l}|\nabla^T\bm{S}_l(\bm{\Phi}_{k,n})\rangle \\+ \langle\bm{S}_l(\bm{\Phi}_{k,n})|\bm{\Lambda}_{z,l}|\nabla\nabla^T\bm{S}_l(\bm{\Phi}_{k,n})\rangle\bigg]\\
    -\langle\tilde{\bm{Z}}_{n,l}|\bm{\Lambda}_{z,l}|\xi_{k,n}\alpha_{k,n,l}\nabla\nabla^T\bm{S}_l(\bm{\Phi}_{k,n})\bigg\}.
\end{multline}
Now carrying out all the expectations we get 
with $\bm{J}_{k,n,l}=\nabla\bm{S}_l(\bm{\Phi}_{k,n})\bigg|_{\bar{\bm{\Phi}}_{k,n}}$
\begin{equation}
    \mathcal{I}_l(\bm{\Phi}_{k,n})\hspace{-1.5pt} =\hspace{-1.5pt} \bar{\xi}_{k,n}(|\bar{\alpha}_{k,n,l}|^2+\sigma_{\alpha,k,n,l}^2)2Re\{\langle\bm{J}_{k,n,l}|\bar{\bm{\Lambda}}_{z,l}|\bm{J}_{k,n,l}\rangle\}.
\end{equation}
For $\bar{\bm{\Lambda}}_{z,l} = \bar{\lambda}_{z,l}\bm{I}$, this simplifies to
\begin{equation}\label{eq:Fisher_information_matrix}
    \mathcal{I}_l(\bm{\Phi}_{k,n}) = \bar{\xi}_{k,n}\bar{\lambda}_{z,l}(|\bar{\alpha}_{k,n,l}|^2+\sigma_{\alpha,k,n,l}^2)2Re\{\langle\bm{J}_{k,n,l}|\bm{J}_{k,n,l}\rangle\}.
\end{equation}
Note that $\bar{\lambda}_{z,l}|\alpha_{k,n,l}|^2\propto SNR$ such that the lowerbound of the variance scales with the \gls{snr} and futher by the variance of the amplitude estimate.  

\renewcommand{\baselinestretch}{0.98}\small\normalsize
\bibliographystyle{IEEEtran}
\bibliography{IEEEabrv,Es_lib}

\end{document}